\def\FullVersion{} 
\def\ShowAuthor{} %
\ifdefined\LLNCS
\documentclass{style/llncs}
\renewcommand{\paragraph}{\subsubsection}
\else
\documentclass[12pt,letterpaper]{article}
\usepackage{amsthm}
\usepackage[margin=1in]{geometry}
\fi
\usepackage{graphicx,color,url,booktabs,comment}
\usepackage{authblk}
\usepackage{amsmath, amssymb,amsbsy}
\usepackage{multicol}
\usepackage{framed}
\usepackage{enumitem}
\usepackage{bbm}
\usepackage{mathrsfs}
\usepackage{braket}
\usepackage{algorithm}
\usepackage{multicol}
\usepackage{algpseudocode}
\usepackage[dvipsnames]{xcolor}
\usepackage{float}
\usepackage{pagecolor}
\usepackage[utf8]{inputenc}
\usepackage{tabularx}
\usepackage[hyperfootnotes=false,,pdfstartview=FitH,colorlinks,linkcolor=blue,filecolor=blue,citecolor=blue,urlcolor=blue]{hyperref}
\usepackage{footnotebackref}
\MakeRobust{\Call}
\makeatletter
\LetLtxMacro{\BHFN@Old@footnotemark}{\@footnotemark}
\renewcommand*{\@footnotemark}{%
    \refstepcounter{BackrefHyperFootnoteCounter}%
    \xdef\BackrefFootnoteTag{bhfn:\theBackrefHyperFootnoteCounter}%
    \label{\BackrefFootnoteTag}%
    \BHFN@Old@footnotemark
}
\makeatother
\usepackage{xspace}
\usepackage{cleveref} 
\usepackage{mdframed}
\usepackage{amsfonts}
\usepackage{tikz}
\usepackage{tikz-cd}
\usepackage{tikz-qtree}
\usepackage{adjustbox}
\usepackage{blindtext}
\usepackage{makecell}
\usepackage{hhline}
\usepackage[noadjust]{cite}
\usepackage{graphicx}
\usepackage[normalem]{ulem}

\usetikzlibrary{arrows,chains,matrix,positioning,scopes,quantikz2}
\ifdefined\LLNCS
\else 
\fi

\setitemize{itemsep=0pt}
\setenumerate{itemsep=0pt}

\newcommand{\customnote}[4]{\ifdefined\Draft{#4\color{#1}[#2: #3]}\fi}
\newcommand{\setnote}[4]{\ifdefined\Draft\newcommand{#1}[1]{\customnote{#3}{#2}{##1}{#4}}\else\newcommand{#1}[1]{}\fi}

\newcommand{\linefill}{\rule{\linewidth}{0.8pt}}

\newcounter{protocol}

\newcounter{securityGame}

\newcounter{balgorithm}
\newenvironment{bAlgorithm}[1]
{
\refstepcounter{balgorithm}
\par\setlength{\parindent}{0pt}
\rule{\textwidth}{0.3mm}
\vspace{-2.6mm}\textbf{Algorithm~\thebalgorithm} #1 

\hrulefill 
}
{
\hrulefill 
\par
}

\newcounter{numConstruction}
\newenvironment{Construction}[1]
{
\refstepcounter{numConstruction}
\par\setlength{\parindent}{0pt}
\rule{\textwidth}{0.3mm}
\vspace{-2.6mm}\textbf{Construction~\thenumConstruction} #1 

\hrulefill 
}
{
\hrulefill \break
\par
}

\newcounter{simulator}

\newenvironment{idealmodel}[1]
{
\begin{mdframed}
    \begin{center}\textbf{#1}\end{center}
}
{
\end{mdframed}
}

\ifdefined\LLNCS
\newenvironment{thm}{\begin{theorem}}{\end{theorem}}

\newenvironment{rmk}{\begin{remark}}{\end{remark}}
\newenvironment{lem}{\begin{lemma}}{\end{lemma}}
\newenvironment{cor}{\begin{corollary}}{\end{corollary}}
\newtheorem{dfn}[definition]{Definition}
\else
\theoremstyle{plain}
\newtheorem{thm}{Theorem}[section]

\newtheorem{rmk}{Remark}

\newtheorem{cor}[thm]{Corollary}

\newtheorem{lemma}[thm]{Lemma}
\newtheorem{claim}[thm]{Claim}

\theoremstyle{definition}
\newtheorem{dfn}[thm]{Definition}
\fi

\ifdefined\LLNCS
\else
\theoremstyle{plain}
\newtheorem*{thm*}{Theorem}
\newenvironment{theorem}{\begin{thm}}{\end{thm}}
\makeatletter
\newtheorem*{rep@theorem}{\rep@title}
\newcommand{\newreptheorem}[2]{%
\newenvironment{rep#1}[1]{%
 \def\rep@title{#2 \ref{##1}}%
 \begin{rep@theorem}}%
 {\end{rep@theorem}}}
\makeatother

\newreptheorem{thm}{Theorem}
\newreptheorem{lem}{Lemma}
\theoremstyle{definition}
\fi

\ifdefined\LLNCS
\newaliascnt{claiml}{theorem}
\newtheorem{claiml}[claiml]{Claim}
\aliascntresetthe{claiml}

\renewenvironment{claim}{\begin{claiml}}{\end{claiml}}

\crefname{claiml}{Claim}{Claims}
\else\fi

\crefname{thm}{Theorem}{Theorems}
\Crefname{thm}{Theorem}{Theorems}
\crefname{lemma}{Lemma}{Lemmas}
\crefname{claim}{Claim}{Claims}
\crefname{figure}{Figure}{Figures}
\crefname{corollary}{Corollary}{Corollaries}
\crefname{proposition}{Proposition}{Propositions}
\crefname{conjecture}{Conjecture}{Conjectures}
\crefname{definition}{Definition}{Definitions}
\crefname{remark}{Remark}{Remarks}
\crefname{example}{Example}{Examples}
\crefname{algorithm}{Algorithm}{Algorithms}
\crefname{bAlgorithm}{Algorithm}{Algorithms}
\crefname{protocol}{Protocol}{Protocols}
\Crefformat{algorithm}{\textup{Algorithm~#2#1#3}}

\renewcommand{\cref}{\Cref} 

\ifdefined\LLNCS
\newenvironment{proofof}[1]{\begin{proof}[of~#1]}{\end{proof}}
\newenvironment{proofsketch}{\begin{trivlist} \item {\it Proof sketch.}} {\qed\end{trivlist}}
\newenvironment{proofsketchof}[1]{\begin{proofsketch}[of~#1]}{\end{proofsketch}}
\else

\newenvironment{proofsketch}{\begin{trivlist} \item {\it Proof sketch.}} {\qed\end{trivlist}}

\fi

\let\mathbb\relax 
\DeclareMathAlphabet{\mathbb}{U}{msb}{m}{n}

\newcommand{\ie}{{i.e.,\ }}
\newcommand{\eg}{{e.g.,\ }}

\def\eps{\epsilon}

\def\cD{\mathcal{D}}

\def\cF{\mathcal{F}}

\def\cK{\mathcal{K}}
\def\cG{\mathcal{G}}

\def\cO{{\cal O}}

\def\cQ{\mathcal{Q}}

\def\cS{{\mathcal{S}}}
\def\cT{{\cal T}}
\def\cU{\mathcal{U}}

\def\sF{\mathsf{F}}

\def\bbC{\mathbb{C}}

\def\bbN{\mathbb{N}}
\def\bbR{\mathbb{R}}

\newcommand{\zo}{\{0,1\}}
\newcommand{\abs}[1]{\left\lvert#1\right\rvert}
\renewcommand{\set}[1]{\left\{#1\right\}}
\newcommand{\norm}[1]{\left\lVert#1\right\rVert}
\newcommand{\op}{\mathsf{op}}

\DeclareMathOperator*{\E}{\mathbb{E}}

\newcommand{\given}{\ensuremath{\;\middle|\;}}
\newcommand{\from}{\leftarrow}

\newcommand{\unif}{\mathsf{unif}}

\DeclareMathOperator{\Sym}{Sym}

\newcommand{\Span}{\mathsf{span}}
\newcommand{\Dom}{\mathsf{Dom}}
\renewcommand{\Im}{\mathsf{Im}}
\newcommand{\intext}{{\operatorname{in}}}
\newcommand{\out}{{\operatorname{out}}}

\newcommand{\secparam}{\lambda}
\newcommand{\negl}{\operatorname{negl}}
\DeclareMathOperator{\poly}{poly}

\newcommand{\adv}{{\mathcal{A}}}

\newcommand{\Gen}{\mathsf{Gen}}

\newcommand{\Eval}{\mathsf{Eval}}

\newcommand{\Sim}{\mathsf{Sim}}

\newcommand{\PRF}{\mathsf{PRF}}
\newcommand{\PRP}{\mathsf{PRP}}
\newcommand{\QObf}{\mathsf{QObf}}
\newcommand{\QEval}{\mathsf{QEval}}

\newcommand{\Den}[1]{\cD^{#1}}
\newcommand{\ketbra}[1]{\ket{#1}\!\bra{#1}}
\newcommand{\identity}{\mathbbm{1}}

\newcommand{\Hg}{\ensuremath{\mathsf{H}}}

\newcommand{\SWAP}{\ensuremath{\mathsf{SWAP}}}
\newcommand{\CTRL}{\mathsf{ctrl}}

\def\ot{\otimes}

\DeclareMathOperator{\Tr}{Tr}

\newcommand{\twirl}{\mathcal{T}}

\def\Ideal{{\mathsf{Ideal}}}
\newcommand{\hybrid}{\mathsf{Hyb}}

\newcommand{\wt}{\widetilde}

\newcommand{\mixednospace}[1]{\mathsf{Mixed}\left[#1\right]}
\newcommand{\mixed}[1]{\mixednospace{#1}}

\newcommand{\Haar}{\mathsf{Haar}}

\newcommand{\Supp}{\mathsf{Supp}}

\newcommand{\PRU}{\mathsf{sPRU}}
\newcommand{\standardPRU}{\mathsf{PRU}}
\newcommand{\spPRU}{\mathsf{spsPRU}}
\newcommand{\arbitrary}{\mathsf{arbitrary}}
\newcommand{\unitary}{\mathsf{unitary}}
\newcommand{\Qarb}{\cQ_{\arbitrary}}

\newcommand{\Quni}{\cQ_{\unitary}}
\newcommand{\Qcla}{\cQ_{\mathsf{pseudo\text{-}det}}}

\newcommand{\frakD}{\mathfrak{D}}

\newcommand{\ztwirl}{\twirl^{(2)}_{{\frakD}_{\mathsf{std}}|_{S^\perp}}}
\newcommand{\xtwirl}{\twirl^{(2)}_{{({\frakD}_{\mathsf{std}} \substack{\circ} H_d)}|_{S^\perp}}}
\newcommand{\threefold}{{\frak{D}_{\mathsf{3fold}}}}
\newcommand{\fulltwirl}{\twirl^{(2)}_{\threefold|_{S^\perp}}}

\newcommand{\bbmI}{\mathbbm{I}}
\newcommand{\bbmF}{\mathbbm{F}}

\newcommand{\sym}{\mathsf{sym}}

\newcommand{\PRO}{\cO_{\mathsf{record}}}

\newcommand{\parameter}{\mathbbm{p}}

\providecommand{\sSym}{\mathsf{Sym}}

\providecommand{\calO}{\mathcal{O}}

\providecommand{\calU}{\mathcal{U}}

\providecommand{\gsA}{{\textcolor{darkgray}{\mathsf{A}}}}
\providecommand{\gsAone}{{\textcolor{darkgray}{\mathsf{A_1}}}}
\providecommand{\gsAtwo}{{\textcolor{darkgray}{\mathsf{A_2}}}}

\providecommand{\gsL}{{\textcolor{darkgray}{\mathsf{L}}}}
\providecommand{\gsLprime}{{\textcolor{darkgray}{\mathsf{L'}}}}

\providecommand{\gsR}{{\textcolor{darkgray}{\mathsf{R}}}}
\providecommand{\gsRprime}{{\textcolor{darkgray}{\mathsf{R'}}}}
\providecommand{\gsS}{{\textcolor{darkgray}{\mathsf{S}}}}

\providecommand{\sF}{\mathsf{F}}

\providecommand{\EPR}{\mathsf{EPR}}
\providecommand{\Id}{\mathsf{Id}}

\providecommand{\eq}{\mathsf{eq}}

\newcommand{\FPC}{{\sf P}_+}
\newcommand{\FPCInv}{{\sf P}_{-}}

\title{Obfuscation of Arbitrary Quantum Circuits\\ \Large
{(via Subspace-Preserving Pseudorandom Unitary)}}
\setnote{\TODO}{TODO}{red}{}
\setnote{\Enote}{Ercheng}{teal}{\footnotesize}
\setnote{\Mnote}{Miryam}{cyan}{\footnotesize}

\ifdefined\ShowAuthor
\author[1]{Mi-Ying (Miryam) Huang\thanks{Email: miying.huang@usc.edu}}\affil[1]{University of Southern California}
\author[2]{Er-Cheng Tang\thanks{Email: erchtang@uw.edu}}\affil[2]{University of Washington}
\date{}

\else
\author{}
\date{}
\fi

\begin{document}
\maketitle

\begin{abstract}
Program obfuscation aims to conceal a computer program's internal structure while preserving its functionality. A central open problem is whether an obfuscation scheme for arbitrary quantum circuits exists. Despite several efforts toward this goal, prior works have succeeded only in obfuscating quantum circuits that implement either pseudo-deterministic functions or unitary transformations. Although unitary transformations already include a broad class of quantum computations, many important quantum tasks, such as state preparation, quantum sampling, and quantum error correction, go beyond unitaries and are described by general completely positive trace-preserving (CPTP) maps.

In this work, we construct the first quantum ideal obfuscation scheme for arbitrary quantum circuits that support quantum inputs and outputs in the classical oracle model assuming post-quantum one-way functions, thereby resolving an open problem raised in Bartusek et al. (STOC 2023), Bartusek, Brakerski, and Vaikuntanathan (STOC 2024), and Huang and Tang (FOCS 2025). At the core of our construction lies a novel primitive that we introduce, called the subspace-preserving strong pseudorandom unitary (spsPRU). An spsPRU is a family of efficient unitaries that fix every vector in a given linear subspace $S$, while acting as a Haar random unitary on its orthogonal complement $S^\perp$ under both forward and inverse oracle queries. Furthermore, by instantiating the classical oracle model with the ideal obfuscation scheme for classical circuits proposed by Jain et al. (CRYPTO 2023) and later enhanced by Bartusek et al. (CRYPTO 2026), our obfuscation scheme can also be realized in the quantumly accessible pseudorandom oracle model.
\end{abstract}

\ifdefined\FullVersion
\newpage
\tableofcontents
\clearpage
\fi

\section{Introduction}
Program obfuscation is a powerful primitive that hides the internal structure of a computer program while preserving its functionality. The idea was first proposed in \cite{diffie1976new} and later formally studied in \cite{barak2001possibility}. In classical cryptography, many cryptographic systems, such as fully homomorphic encryption, functional encryption, and universal computational extractor, can be built from obfuscation \cite{ishai2000randomizing,garg2016candidate,brzuska2014indistinguishability,jain2021indistinguishability,jain2022indistinguishability}. 

Given the power of obfuscation in classical cryptography, a natural question is whether program obfuscation for general quantum computation exists in the quantum world.
The strongest notion of obfuscation, known as ideal obfuscation, has been shown to be impossible for general circuits in the plain model under standard cryptographic assumptions \cite{barak2001possibility,alagic2021impossibility}. Consequently, researchers have shifted toward weaker notions of obfuscation, such as quantum indistinguishability obfuscation (iO) \cite{broadbent2021constructions,bartusek2022indistinguishability}, or toward constructions in stronger models \cite{bartusek2022indistinguishability,bartusek2023obfuscation,coladangelo2024use,bartusek2024quantum,huang2025obfuscation}. In the direction of weaker notions, \cite{broadbent2021constructions} constructed a quantum iO scheme for circuits with logarithmically many non-Clifford gates, while \cite{bartusek2022indistinguishability} proposed a quantum iO scheme for circuits where the output is always zero. However, these constructions are limited to highly restricted classes of circuits.
To support broader functionalities, subsequent works have considered stronger models, such as the classical oracle model. In this setting, the works of \cite{bartusek2023obfuscation,bartusek2024quantum,huang2025obfuscation} achieved the strong notion of ideal obfuscation for pseudo-deterministic or unitary quantum computations. Despite these advances, existing quantum obfuscation schemes ultimately only support functionalities that fall within the scope of unitary transformations.

However, unitary transformations do not capture the full generality of quantum computation. There is still a significant gap between the class of unitary quantum circuits and the class of general quantum circuits. For example, the computation of isometries (\eg state preparation circuits), the computation of randomized functions (\eg quantum sampling circuits), and other general quantum computation (\eg quantum error correction circuits) are not unitary transformations.
Consequently, none of the existing obfuscation schemes can obfuscate these fundamental classes of non-unitary quantum computations. 
Furthermore, many quantum cryptographic primitives (\eg quantum money \cite{aaronson2012quantum}, pseudorandom states \cite{ji2018pseudorandom}, and EFI pairs \cite{brakerski2022computational}) involve non-unitary quantum circuits, and therefore failing to obfuscate beyond unitary circuits might hinder further applications of obfuscation to quantum cryptography. This leads us to the following question:

\begin{quote}
    \centering
    \it Is it possible to obfuscate arbitrary quantum circuits?
\end{quote}

\subsection{Our Result}
We answer the above question affirmatively and thus resolve an open problem raised in \cite{bartusek2023obfuscation,bartusek2024quantum,huang2025obfuscation} by constructing the first full-fledged quantum obfuscation scheme for \emph{arbitrary} quantum circuits in the classical oracle model, where the circuits considered are classically described and can implement arbitrary quantum maps with quantum inputs and outputs. Moreover, our scheme achieves ideal obfuscation. To formalize this notion, we develop a unified definitional framework\footnote{Previously, ideal obfuscation was only defined when the circuit computes a deterministic function or a unitary transformation, and the two definitions have different syntax.} that is independent of the specific type of functionality being computed (\cref{def:ideal-obfuscation}). We then obtain a concrete definition of ideal obfuscation by instantiating the framework via an exact characterization of what a circuit implementation of any given quantum map would generically reveal (\cref{fig:our-ideal-functionality}).

\begin{thm}[Quantum Ideal Obfuscation, Informal]
There exists a quantum obfuscation scheme for arbitrary quantum circuits supporting quantum inputs and outputs, achieving the security notion of ideal obfuscation (Definition \ref{def:ideal-obfuscation}) in the classical oracle model assuming post-quantum one-way functions. 
\end{thm}

Our obfuscation is secure in the classical oracle model, in which the obfuscator outputs an efficient classical oracle that is accessible to all parties, and every party can make quantum queries to the oracle. Such a model is equivalent to assuming post-quantum ideal obfuscation for all classical circuits. Although ideal obfuscation for general circuits is impossible \emph{in the plain model} \cite{barak2001possibility}, Jain et al. \cite{jain2023pseudorandom} introduced classical ideal obfuscation for general circuits \emph{in the pseudorandom oracle model}, where only idealized hash functions are assumed. More recently, Bartusek et al. \cite{bartusek2026new} extended this line of work by proving its post-quantum security, where a quantum adversary can query the pseudorandom oracle in superposition. The classical oracle in our obfuscation can be instantiated using post-quantum ideal obfuscation. This allows our construction to operate fully in the pseudorandom oracle model and leads to the following corollary.

\begin{cor}[Obfuscation in the Pseudorandom Oracle Model]
There exists a quantum obfuscation scheme for arbitrary quantum circuits, achieving the security notion of ideal obfuscation (\cref{def:ideal-obfuscation}) in the (quantumly accessible) pseudorandom oracle model assuming a post-quantum subexponential-secure one-way function and functional encryption.
\end{cor}

A key ingredient underlying our full-fledged quantum obfuscation construction 
is the introduction of \emph{subspace-preserving strong pseudorandom unitaries} (spsPRUs), 
which may be of independent interest. Unlike standard strong pseudorandom 
unitaries (sPRUs) on $n$ qubits \cite{ji2018pseudorandom,ma2025construct}, spsPRUs 
simultaneously enforce exact structure on a designated subspace while maintaining 
pseudorandomness on its orthogonal complement.
This additional structure is indispensable when applied to our obfuscation construction. In particular, 
our approach requires enforcing exact behavior on part of the Hilbert space while 
hiding information on the rest, a feature that cannot be achieved by existing notions 
of sPRUs. 

Our construction of spsPRUs departs from the unitary ensemble considered in \cite{ma2025construct}, whose ensemble is inherently limited to prime-power dimensions. Their work critically relied on the fact that the Clifford group forms a unitary $2$-design, a property that fails in general dimensions \cite{graydon2021clifford}. We overcome this barrier by developing a subspace-preserving approximate restricted $2$-design together with a careful analysis of its properties that hold in any dimension. Moreover, to the best of our knowledge, this work provides one of the first applications of (a variant of) PRU that relies on all properties of PRU.\footnote{PRU implies pseudorandom isometries (PRI) and pseudorandom function-like states (PRFS), and most existing applications of PRU can be based solely on PRI or PRFS. However, for our obfuscation construction, PRI or PRFS alone do not seem to suffice.}

\begin{thm}[Subspace-Preserving Strong Pseudorandom Unitary (spsPRU), Informal]
There exists an $S$-preserving strong pseudorandom 
unitary (\cref{def:spsPRU}) for some choices of the linear subspace 
$S$, assuming post-quantum one-way functions.
\end{thm}

\begin{rmk} \label{rmk:generictrans}
The spsPRU plays an important role in our obfuscation. Although our scheme is presented as a concrete construction, it is in fact a generic transformation that compiles an underlying obfuscation scheme into one for arbitrary quantum circuits. In this sense, the spsPRU enables a black-box lifting and eliminates the need for ad-hoc constructions. \end{rmk}

\paragraph{Related Work}
A central goal in quantum obfuscation is to achieve obfuscation for 
\emph{classically described quantum circuits} 
\cite{broadbent2021constructions,bartusek2022indistinguishability,bartusek2023obfuscation,bartusek2025classical}, 
closely mirroring classical obfuscation for classical circuits. Progress toward this goal has been gradual. Prior works have established obfuscation 
for restricted classes of circuits, including those with $\log$-many non-Clifford gates 
\cite{broadbent2021constructions}, null circuits \cite{bartusek2022indistinguishability}, and pseudo-deterministic circuits \cite{bartusek2023obfuscation,bartusek2025classical}. At the same time, a complementary line of work considers another form in which 
circuits may carry auxiliary quantum states, leading to the notion of obfuscating 
\emph{quantumly described quantum circuits} \cite{coladangelo2024use,bartusek2024quantum,huang2025obfuscation}. 
This perspective enables applications such as best-possible copy protection\footnote{It is worth noting that achieving best-possible copy protection through quantum obfuscation for quantumly described quantum circuits does not preclude obtaining it via quantum obfuscation for classically described quantum circuits, as whether $\mathsf{BQP/poly} \subsetneq \mathsf{BQP/qpoly}$ remains open.} 
\cite{coladangelo2024use}, and has led to constructions for broader classes of functionalities, 
including unitary transformations \cite{huang2025obfuscation}. Despite this progress on obfuscating quantumly described quantum circuits, obfuscating classically described quantum circuits remains the more natural and central goal, as quantum computations are typically specified in a classically described form.

\begin{table}[!ht]
\renewcommand{\arraystretch}{1.8}
\scriptsize
\centering
\resizebox{\textwidth}{!}{
\begin{tabular}{|c|c|c|c|c|c|c|}
    \hline
    {} & \makecell{Obfuscator \\ input} & \makecell{Obfuscator \\ output} & \makecell{Program \\ input} & \makecell{Program \\ output} & Program class & Result \\
    \hhline{|=|=|=|=|=|=|=|}
    \cite{broadbent2021constructions} & Classical & Quantum & Quantum & Quantum & \makecell{Unitary circuits with \\log-many non-Clifford gates}  & iO\\
    \hline
    \cite{bartusek2022indistinguishability} & Classical & Classical & Quantum & Classical & Null circuits & iO \\
    \hline
    \cite{coladangelo2024use} & Quantum & Quantum & Classical & Classical & Deterministic circuits & iO \\
    \hline 
    \cite{bartusek2023obfuscation} & Classical & Quantum & Classical & Classical & \makecell{Pseudo-deterministic circuits} & VBB \\
    
    \hline
    \cite{bartusek2024quantum} & Quantum & Quantum & Classical & Classical & \makecell{Pseudo-deterministic circuits} & Ideal\\
    
    \hline
    \cite{bartusek2025classical} & Classical & Classical & Classical & Classical & \makecell{Pseudo-deterministic circuits} & Ideal\\
    
    \hline
    \cite{huang2025obfuscation} & Quantum & Quantum & Quantum & Quantum & \makecell{Approximately-unitary circuits}  & Ideal\\
    
    \hline
    
    This work & Classical & Quantum & Quantum & Quantum & \makecell{Arbitrary circuits} & Ideal\\
    \hline
\end{tabular}
}
\caption{Constructions of quantum obfuscation schemes.}
\label{table:sum}
\end{table}

We summarize the comparison between our result and prior quantum obfuscation constructions in Table~\ref{table:sum}. Here, obfuscator input indicates whether the obfuscator takes classically described quantum circuits or quantumly described quantum circuits (\ie with an auxiliary quantum state). Obfuscator output specifies whether the obfuscated program contains quantum states.


\paragraph{Potential Applications.} 
In the work of \cite{bartusek2023obfuscation}, quantum obfuscation was employed to construct a quantum functional encryption scheme and quantum copy-protection schemes. By instantiating their construction with our obfuscator for quantum computations that produce possibly randomized classical outputs, one might obtain a functional encryption scheme supporting general functionalities. However, since the outputs are no longer required to be deterministic, the security analysis could be non-trivial. We leave this as an open question for those interested in further exploration.

Our obfuscation may also be a helpful tool for AI security. Since the computational functionality of arbitrary quantum circuits covers all classical randomized functions (whereas classical deterministic functions or unitary transformations do not subsume classical randomized functions), if, perhaps in the distant future, large language models (LLMs) are enhanced using quantum computation, our construction could provide a natural candidate for watermarking or obfuscating such programs.

\paragraph{Open Problems}
\begin{enumerate}
    \item Our work constructs quantum ideal obfuscation in the oracle model. It has been shown that ideal obfuscation is not achievable in the plain model \cite{barak2001possibility,alagic2021impossibility}. This naturally leads to the next question: whether it is possible to construct quantum indistinguishability obfuscation (QiO) in the plain model under standard assumptions.
    \item Classically, iO together with one-way functions implies almost all of cryptography. In the quantum setting, a natural question is whether QiO for arbitrary quantum circuits, combined with post-quantum one-way functions, pseudorandom states, or quantum commitments, can imply comparable primitives in quantum cryptography.
    \item Subspace-preserving strong pseudorandom unitaries (spsPRUs) play a core role in constructing our quantum obfuscation, and it would be interesting to explore whether additional applications can be built upon them.

\end{enumerate}

\subsection{Technical Overview}
\label{sec:overview}

In this section, we first review the state of the art in quantum obfuscation and explain why existing techniques cannot obfuscate arbitrary quantum circuits. We then present the high-level idea of our solution based on a novel primitive that we introduce, which we call a subspace-preserving strong pseudorandom unitary (spsPRU), and describe the resulting obfuscation construction. Next, we discuss the difficulty of formalizing the ideal functionality of obfuscation for arbitrary quantum circuits and outline how we overcome these challenges. We conclude the section with a proof sketch establishing that our construction achieves ideal obfuscation for arbitrary quantum circuits, and how spsPRU can be constructed.

\subsubsection{Challenge of Obfuscating Arbitrary Quantum Circuits}
\label{sec:challenge:obf}
There are two main approaches to constructing quantum obfuscations. The first approach \cite{bartusek2023obfuscation,bartusek2025classical} employs quantum fully homomorphic encryption (QFHE) and classical verification of quantum computation (CVQC) \cite{mahadev2018classical}. However, since the verifier in CVQC protocols can only verify classically described statements, this methodology is fundamentally restricted to handling quantum circuits whose inputs and outputs are classical.
Moreover, existing CVQC schemes with negligible error are restricted to decision problems; in contrast, all known CVQC schemes for sampling problems suffer from inverse-polynomial error \cite{chung2022constant}. Therefore, any approach along this line is inherently subject to pseudo-deterministic limitations and thus remains far from achieving obfuscation for arbitrary quantum circuits.

The second approach \cite{bartusek2024quantum,huang2025obfuscation} is based on quantum authentication schemes that support homomorphic computation, where the obfuscated circuit would carry an auxiliary authenticated quantum state that encodes the original circuit but hides the circuit information. This line of work has achieved the obfuscation of quantum circuits implementing unitary transformations \cite{huang2025obfuscation}. Nevertheless, when attempting to obfuscate arbitrary quantum circuits computing general quantum maps (\ie completely positive trace-preserving maps), it would be challenging to enable the reusability of the authenticated quantum state, which may be disturbed after each evaluation. The work of \cite{huang2025obfuscation} critically restricts attention to unitary quantum circuits because for unitary functionalities, the output is guaranteed to be disentangled from the auxiliary state after execution, allowing the auxiliary state to be coherently restored (e.g., via rewinding) and reused in subsequent evaluations. This structural guarantee fails for non-unitary programs, as the auxiliary state may remain entangled with the output or even become part of it, leaving no generic method to restore the original auxiliary state without disturbing the computation. Additionally, entanglement between the adversary’s output states and the auxiliary states of the obfuscated program can complicate the security analysis of the obfuscated program's behavior. As a result, existing techniques cannot support obfuscation beyond the computation of unitary transformations.

Consider a quantum circuit that maps an $n$-qubit input to $n' > n$ qubits of output. Any description of such a circuit must involve the preparation of ancilla qubits, and some of the ancilla qubits must become part of the output. Ensuring that these ancilla qubits are properly prepared is crucial and introduces additional challenges. A natural attempt is to let the obfuscator prepare the required ancilla qubits and embed them in the auxiliary state of the obfuscated program. As some of the ancilla qubits are irreversibly consumed during the circuit's evaluation, the resulting construction is intrinsically non-reusable. Thus, it seems more viable to instead let the evaluator prepare and supply the required ancilla qubits during execution. Yet this alternative also introduces challenges: the obfuscated program must verify that the supplied ancillas are correctly prepared before running the computation. Attempting to perform such verification by conditionally applying the computation controlled on the ancillas being correctly prepared is unfortunately not a valid quantum operation, since both the control and the computation act on the same ancilla registers.
Taken together, the reusability and ancilla verification issues obstruct the generalization of existing obfuscation techniques to arbitrary quantum functionalities.

\subsubsection{Our Solution and Construction}

Our approach for obfuscating arbitrary quantum circuits involves transforming a given circuit $Q$ that implements a general quantum map $\Phi$ into a \emph{unitary circuit} $Q'$ that both hides $Q$ and enables the evaluation of $\Phi$.
This transformation is enabled by a novel primitive that we introduce, called a \emph{subspace-preserving strong pseudorandom unitary} (spsPRU). We then apply an obfuscation scheme for unitary circuits\footnote{As mentioned earlier in \cref{rmk:generictrans}, our obfuscation transformation imposes no restrictions on the choice of ideal obfuscation scheme for unitary quantum circuits. In this work, we instantiate this component using the scheme of \cite{huang2025obfuscation}. However, any construction for ideally obfuscating all unitary functionalities can be plugged into our generic transformation to yield an obfuscation scheme for arbitrary quantum circuits.} as a building block to obtain the final construction.

Our construction begins with purifying any given quantum circuit $Q$ via the standard technique of deferring all measurements in $Q$ to the end of the circuit. This yields an equivalent circuit that takes a quantum input, prepares ancilla qubits in the $\ket{1}$ state, applies a sequence of unitary gates to both the input and ancilla registers, and outputs a subset of the qubits. We then form a unitary circuit $U_Q$ consisting only of this sequence of unitary gates, which acts jointly on the input and ancilla registers.

However, the circuit $U_Q$ itself does not impose any restriction that the ancillas must all be in the $\ket{1}$ state. To resist attacks that compute using incorrect ancilla states, subspace-preserving strong pseudorandom unitaries come to the rescue. In more detail, an $S$-preserving strong PRU is a family $\set{\spPRU_k}_{k}$ of efficient unitaries that fixes every element in a given linear subspace $S$ while acting like a Haar random unitary on the orthogonal subspace $S^\perp$ in the sense that the family is indistinguishable from the Haar measure over $\set{U \in \cU(2^n) : U \ket{v} = \ket{v} \text{ for all } \ket{v} \in S \subseteq \bbC^{2^n}}$ under forward and inverse queries.
Compared to the ordinary strong PRU \cite{ma2025construct}, our subspace-preserving strong PRU exhibits both structure (in $S$) and pseudorandomness (in $S^\perp$), where $S$ can be chosen to have any dimension.
By setting $S$ as the subspace of states with correctly prepared ancillas, one can use an $S$-preserving strong PRU to ensure that the input and ancillas remain in the same state, conditioned on the ancillas all being in the $\ket{1}$ state, while fully scrambling the inputs and ancillas into a pseudorandom state (in $S^\perp$), conditioned on the ancillas being incorrectly prepared.

Together, we form a unitary circuit $Q' = (\PRU_{k'} \ot I) \circ U_Q \circ \spPRU_k$ for random $k, k'$, where $\PRU$ is an ordinary strong PRU that acts on the qubits not being output by $Q$. 
By design, $\spPRU$ does not affect the computation of $Q$ when the ancillas are honestly prepared. Therefore, $Q'$ performs the same computation as $Q$, provided that the ancillas are correctly initialized and only the designated output qubits are obtained. Then, we obfuscate the unitary circuit $Q'$ using the obfuscator of \cite{huang2025obfuscation} to produce the obfuscated program. The final adjustment is to select the orthogonal subspace  $S^\perp$ with sufficiently large dimension so that the resulting security loss remains negligible. Taking this into account yields our obfuscation construction as follows.

\paragraph{Our Obfuscation Construction.}
Given a security parameter $\secparam$ and a quantum circuit $Q$ with $n$ input qubits and $m \ge n$ total working qubits, our obfuscator performs the following.
\begin{enumerate}
    \item Increase the number of working qubits to $m'=\lambda+m$ qubits and set $d=2^{m'}-2^n$.
    \item Set the subspace $S = \Span\{\ket{1^{m'-n}, y}\}_{y\in \zo^{n}} = \Span\{\ket{x}\}_{x\in \zo^{m'} \setminus [d]}$.\footnote{Here, we identify numbers in $[d] = \set{0,1,\dots,d-1}$ with their binary representations in $\zo^{m'}$.}\\
    Therefore, $S^\perp = \Span\{\ket{x}\}_{x\in [d]}$ has dimension $d = \Omega(2^\secparam)$.
    \item Form the unitary circuit $Q'=(\PRU_{k'}\ot I)\circ (\CTRL_{1^\lambda}\text{-}U_Q)\circ \mathsf{spsPRU}_k$ for random $k,k'$, where $\mathsf{spsPRU}$ is an $S$-preserving strong PRU. 
    \item Output the obfuscated program produced by applying the obfuscator of \cite{huang2025obfuscation} to the unitary circuit $Q'$.
\end{enumerate}

\subsubsection{Ideal Obfuscation: Definition and Security Analysis}

Ideal obfuscation, typically defined through a simulation-based formulation, is a stronger notion than indistinguishability obfuscation and serves as a more conceptually ambitious goal when studying obfuscation in idealized models. In particular, ideal obfuscation requires that the obfuscated circuit be efficiently simulatable from black-box access to the input–output behavior of the original circuit and its size. This formulation provides a direct characterization of the information available from the obfuscated circuit, making it explicit that the obfuscated circuit reveals nothing dependent on the original implementation details.

Nevertheless, existing formulations of ideal obfuscation apply only to the computation of either a deterministic classical function $F$ or a unitary transformation 
$\cU$. In the unitary case, choosing an appropriate notion of the input-output behavior already requires some care, as the simulator needs query access to both $\cU$ and $\cU^{-1}$ to account for the fact that any implementation of $\cU$ inherently allows one to also compute its inverse $\cU^{-1}$  \cite{huang2025obfuscation}.
Defining ideal obfuscation in full generality, \ie for the computation of a general quantum mapping $\Phi$, appears even more challenging because unlike the unitary case where invertibility guides the definition, a general quantum map $\Phi$ admits no canonical structure to rely on. We address the aforementioned issues in the following two steps.

First, we reformulate the definition of ideal obfuscation to make the definition independent of the type of functionality being computed, while ensuring that the reformulated definition accurately captures the semantic meaning of ideal obfuscation. This is achieved by augmenting the definition with the idea of an ``ideal functionality,'' which plays the role of representing the input-output behavior of any quantum circuit computing $\Phi$ (together with the circuit size). In other words, our ideal obfuscation definition now requires that the obfuscated circuit can be efficiently simulated by a simulator that interacts with an ideal functionality corresponding to $\Phi$. Additionally, we require that the choice of ideal functionality has to faithfully represent what can be computed from any quantum circuit that implements $\Phi$. In particular, it must not provide additional computational power beyond what can be generically extracted from a circuit implementing $\Phi$, since an overly powerful ideal functionality, \eg one internally equipped with a PSPACE oracle, would render the resulting security weak.

Second, we identify a suitable ideal functionality for a general quantum mapping $\Phi$ by establishing the existence and uniqueness of an intrinsic distribution $\mu^\unif_\Phi$ of unitary operators that can be used to directly compute $\Phi$. This distribution depends only on the quantum mapping $\Phi$ and the amount of space required for the implementation. 
Building on this, we define the corresponding ideal functionality as giving out the security parameter and circuit parameters (\eg size), and providing forward and inverse queries to a unitary sampled from the distribution $\mu^\unif_\Phi$. 
Crucially, this definition of ideal functionality is independent of any specific implementation of $\Phi$, besides the circuit parameters themselves.
Furthermore, we establish that our proposed ideal functionality does not contain more capabilities than needed, by showing that the ideal functionality can always be efficiently implemented given the security parameter and any circuit realizing $\Phi$ with the prescribed parameters, thereby demonstrating that the proposed ideal functionality precisely captures the computational power of white-box access to any such circuit implementation of $\Phi$.

Finally, we briefly discuss the high-level idea of the security analysis. First, by the security of the ideal obfuscation scheme for unitary circuits, our obfuscated program can be simulated using forward and inverse queries to $$Q' =(\PRU_{k'}\ot I)\circ (\CTRL_{1^\lambda}\text{-}U_Q)\circ \spPRU_k.$$
Then, we apply the pseudorandomness of $\PRU,\spPRU$ to replace $\PRU$ with a Haar random unitary and $\spPRU$ with a unitary that acts as the identity on $S$ and is Haar random on $S^\perp$. The resulting modified unitary $Q'$ exactly follows the distribution $\mu^{\unif}_\Phi$ discussed above, implying that our obfuscated program can be simulated using the ideal functionality associated with $\Phi$. Therefore, our obfuscation scheme achieves ideal obfuscation security.
Moreover, we show that the distributions $\mu^\unif_\Phi, \mu^\unif_{\Phi'}$ are statistically close whenever $\Phi,\Phi'$ are close in diamond distance, and thus our ideal obfuscation scheme also satisfies the notion of indistinguishability obfuscation. We defer the formal details and proofs to \cref{sec:obfuscation}.

\subsubsection{Subspace-Preserving Strong PRU}
\label{sec:overview-spsPRU}

As discussed earlier, our obfuscation construction makes heavy use of a primitive we introduce, called a subspace-preserving strong pseudorandom unitary (spsPRU). In contrast to the standard notion of strong pseudorandom
unitaries (sPRU) on $n$ qubits \cite{ji2018pseudorandom,ma2025construct}, spsPRU requires enforcing exact behavior on a designated subspace $S$ while remaining pseudorandom on its orthogonal complement $S^\perp$. This additional structure makes spsPRU more delicate than sPRU. A natural construction approach would be to adapt the unitary ensemble from \cite{ma2025construct}, but their ensemble critically relies on the Clifford group forming a unitary 2-design, a property that fails when the dimension is not a power of a prime \cite{graydon2021clifford}. We address the limitation and obtain subspace-preserving strong PRUs in which pseudorandomness can be achieved on subspaces of general dimensions.

\paragraph{Constructing spsPRU}
We now describe how we construct a subspace-preserving PRU for a subspace of the form $S = \Span\{\ket{x}\}_{x\in\{0,1\}^n\setminus[d]}$ with $d \le 2^n$. In other words, we want a family $\set{\spPRU_k}_k$ of efficient unitaries that preserves every vector $\ket{x}$ for $x \in\{0,1\}^n\setminus[d]$, and is pseudorandom on $S^\perp=\Span\{\ket{x}\}_{x\in[d]}$. Our construction is inspired by the unitary ensemble $D \cdot P_\pi \cdot F_f \cdot C$ considered in \cite{ma2025construct}, in which $C,D$ are drawn from an exact unitary $2$-design $\frakD$ instantiated with the uniform distribution over the Clifford group, $P_\pi$ permutes the computational basis states according to a pseudorandom permutation $\pi$, and $F_f$ applies phases to the computational basis states according to a pseudorandom function $f$. For the purpose of constructing an $S$-preserving strong PRU, however, analogous ensembles do not work, as we need pseudorandomness on the $d$-dimensional subspace $S^\perp$, but the $d$-dimensional Clifford group is generally not a unitary $2$-design \cite{graydon2021clifford}. To resolve this issue, we show that the requirement of $\frakD$ being an exact unitary $2$-design can be relaxed to a notion of approximate restricted $2$-design that is even weaker than being an approximate $2$-design. Furthermore, we draw inspiration from \cite{nakata2017unitary} to construct an $S$-preserving variant of an approximate restricted $2$-design for the subspace $S = \Span\set{\ket{x}}_{x \in \zo^n \setminus [d]}$ using the unitary ensemble $Z_{f_3} \cdot H_d \cdot Z_{f_2} \cdot H_d \cdot Z_{f_1}$, where $H_d$ is the Fourier transform on $\Span\set{\ket{x}}_{x\in[d]}$ and $Z_{f_1},Z_{f_2},Z_{f_3}$ apply phases to the computational basis states $\set{\ket{x}}_{x\in[d]}$ according to random functions $f_1,f_2,f_3$. Together, by combining the $S$-preserving approximate restricted $2$-design with existing works on pseudorandom permutations that fix elements in $\{0,1\}^n\backslash[d]$ and are pseudorandom over $[d]$, we obtain an $S$-preserving strong PRU. The formal details can be found in \cref{Sec:PRU}.

\subsection{Roadmap}
Section \ref{Sec:prelim} prepares readers with some preliminaries. 
Section \ref{Sec:PRU} constructs a subspace-preserving strong pseudorandom unitary. 
Section \ref{sec:obfuscation} presents a quantum obfuscation scheme for arbitrary quantum circuits  achieving ideal obfuscation.

\section{Preliminary}
\label{Sec:prelim}
Sampling uniformly from a set $S$ is denoted by $s\from S$.
A function $f \colon \bbN \to [0,1]$ is called negligible if for every polynomial $\poly(\cdot)$, it holds that $f(n)<|{1}/{\poly(n)}|$ for all sufficiently large $n$. We use $\negl(\cdot)$ to denote an unspecified negligible function. QPT stands for quantum polynomial time.

An $n$-qubit pure state $\ket{\phi}$ is a unit vector in the Hilbert space $\bbC^{2^n}$, and is identified with the density matrix $\mixed{\ket{\phi}} = \phi = \ketbra{\phi} \in \bbC^{2^n \times 2^n}$. The set of $n$-qubit (mixed) states, denoted $\Den{n}$, consists of positive semi-definite matrices in $\bbC^{2^n \times 2^n}$ with trace $1$.
A state $\rho$ on a register $\gsR$ is denoted as $\rho_{\gsR}$.
The notation $(\rho, \sigma)$ denotes a possibly entangled state on two registers. Let $\Tr$ denote the trace. For a unitary operator $U$ on $\bbC^{2^n}$, its controlled unitary operator on $\bbC^{2} \ot \bbC^{2^n}$ is denoted as $\CTRL\text{-}U := \ketbra{0} \ot I + \ketbra{1} \ot U$.

A quantum map $\mathcal{F}: \Den{n} \to \Den{n'}$ is a completely positive trace-preserving (CPTP) map. The quantum map of applying an operator $V$ refers to the map $\rho \mapsto V \rho V^\dag$.
The partial trace $\Tr_{\gsR}$ over register $\gsR$ is the unique linear map from registers $\gsR,\gsS$ to $\gsS$ such that $\Tr_{\gsR}(\rho_{\gsR} \ot \sigma_{\gsS}) = \Tr(\rho) \sigma$ for every product state $\rho \ot \sigma$. We write $\Tr_1$ as the partial trace over the first register. For multi-qubit systems, we write $\Tr_{[:-n]}$ as the partial trace over all but the last $n$ qubits.

\subsection{Quantum Circuit and Distance Measure}
\label{sec:quantum-circuit-and-norm-distance}
A quantum circuit $Q$ is described by a sequence of unitary gates and standard-basis measurements, followed by a specification of the output bits or qubits. The width of $Q$ is the number of qubits it acts on, and the size of $Q$ is the number of gates and measurements it has.

The operator norm of a linear operator $ A $ is defined as $\norm{A}_{\mathsf{op}} := \max_{\ket{v}} \frac{\norm{Av}}{\norm{v}}$ and the trace norm  of $A$ is defined as $\norm{A}_1 := \Tr(\sqrt{A^\dagger A})$. The operator norm of a linear map $\cF:\Den{n} \to \Den{n'}$ is defined as $\norm{\cF}_{\mathsf{op}} := \max_{\rho \in \Den{n}} \norm{\cF(\rho)}_1$. For quantum maps $\cF,\cG: \Den{n} \to \Den{n'}$, their diamond distance is defined as
$\| \cF- \cG \|_\diamond := \max_{\rho \in \Den{2n}} \| (\cF \otimes \identity_n) (\rho) - (\cG \otimes \identity_n) (\rho) \|_{1}$.

\begin{theorem}[\cite{kretschmann2008information}] \label{thm:continuity-of-dilation}
    Let $\Phi_0, \Phi_1: \Den{n} \to \Den{n'}$ be quantum maps and $V_0, V_1: \bbC^{2^n} \to \bbC^{2^k} \ot \bbC^{2^{n'}}$ be isometries such that
    $\Phi_i(\rho) \equiv \Tr_{1} ( V_i \rho V_i^\dag )$ for $i=0,1$. Then we have
    $$ \inf_{U \in \cU(2^k)} \norm{(U \ot I) V_0 - V_1}_{\mathsf{op}} \le \sqrt{\norm{\Phi_0 - \Phi_1}_\diamond} $$
\end{theorem}

\begin{lemma}[\cite{jordan1875essai}]
\label{lem:Jordan}
Let $\Pi_A,\Pi_B$ be projectors on a Hilbert space $\mathcal{H}$. Then $\mathcal{H}$ can be decomposed as a direct sum of one-dimensional and two-dimensional subspaces $\{\cS_j\}_j$ (the Jordan spaces), $\mathcal{H} = \bigoplus_j \mathcal{S}_j$, where each $\mathcal{S}_j$ is invariant under both $\Pi_A$ and $\Pi_B$. Moreover,
\begin{itemize}
    \item[-] In every one-dimensional subspace, $\Pi_A$ and $\Pi_B$ act either as the identity or as the zero operator.
    \item[-] In every two-dimensional subspace $\mathcal{S}_j$, both $\Pi_A$ and $\Pi_B$ are rank-one projectors. 
    In this case, there exist orthonormal bases 
    $\{ \ket{v_{j,1}}, \ket{v_{j,0}} \}$ and $\{ \ket{w_{j,1}}, \ket{w_{j,0}} \}$ 
    of $\mathcal{S}_j$ such that 
    $\Pi_A$ projects onto $\ket{v_{j,1}}$ and $\Pi_B$ projects onto $\ket{w_{j,1}}$.
\end{itemize}
\end{lemma}

\subsection{Topological Group and Haar Measure}
\label{sec:topo-haar}

The $d$-dimensional unitary group is denoted as $\cU(d)$. Let $d \le N = 2^n$. We identify each integer $x$ in $[N] = \set{0,1,\dots,N-1}$ with its binary representation in $\zo^n$. Given a $d$-dimensional subspace $S_0 \subseteq \bbC^{2^n}$, we define the topological group $$\cU(S_0) := \set{U \in \cU(2^n) : U \ket{v} = \ket{v} \text{ for all } \ket{v} \in S_0^\perp}$$
as the subgroup of $\cU(2^n)$ that acts trivially on $S_0^\perp$. The group $\cU(S_0)$ is isomorphic to $\cU(d)$. 

\begin{dfn}[Haar measure]\label{def:haar}
The Haar measure $\mu^{\Haar}_{S_0}$ on the (compact) topological group $\calU(S_0)$ is the unique probability measure on the Borel measurable subsets of $\calU(S_0)$ that is both left and right invariant: For every Borel measurable set $A \subseteq \calU(S_0)$ and 
every $U \in \calU(S_0)$, the probability measure $\mu^{\Haar}_{S_0}$ satisfies $$\mu^{\Haar}_{S_0}(UA) = \mu^{\Haar}_{S_0}(AU) = \mu^{\Haar}_{S_0}(A)$$
The Haar measure on $\cU(d)$ is denoted as $\mu^\Haar_d$. We identify the measure $\mu^\Haar_{S_0}$ on $\cU(S_0)$ with its pushforward measure on $\cU(2^n)$ under the inclusion map $\cU(S_0) \xhookrightarrow{} \cU(2^n)$.
\end{dfn}

\begin{dfn}[Pushforward measure]
Let $(X, \sigma_X)$ and $(Y, \sigma_Y)$ be measurable spaces, $\mu$ be a measure on $(X,\sigma_X)$,
and $f : X \to Y$ be a measurable function. 
The pushforward measure of $\mu$ under $f$, denoted by $f_* \mu$, 
is the unique measure on $(Y, \sigma_Y)$ satisfying

$$\int_Y g(y) \, (f_* \mu)(dy)
:= 
\int_X g(f(x)) \, \mu(dx)
\quad
\text{for all measurable } g : Y \to \mathbb{R}.$$
\end{dfn}

\begin{dfn}[$t$-wise twirl] The $t$-wise twirl with respect to a distribution $\frakD$ over $\cU(d)$ is defined as the operator
$$\cT^{(t)}_{\frakD}(\cdot) \coloneqq \E_{U \sim \frakD} U^{\ot t} (\cdot) U^{\dag, \ot t}$$
\end{dfn}

\begin{dfn}[$t$-design]
A distribution $\frakD$ over $\cU(d)$ is a unitary $t$-design if
$$\cT_{\frakD}^{(t)}(\cdot) = \cT_{\mu^\Haar_d}^{(t)}(\cdot)$$
\end{dfn}

We consider the following operators on $\bbC^{d} \ot \bbC^{d}$.
\begin{itemize}
    \item $\Pi^\eq \coloneqq \sum_{x \in [d]} \ketbra{x,x}$.
    \item $\Pi^\sym \coloneqq \Pi^{\eq} + \sum_{x > y} \left(\frac{\ket{x,y} + \ket{y,x}}{\sqrt{2}}\right) \left( \frac{\bra{x,y} + \bra{y,x}}{\sqrt{2}}\right)$ projects onto the symmetric subspace.
    \item $\Pi^\EPR \coloneqq \left(\frac{1}{\sqrt{d}}\sum_{x \in [d]} \ket{x,x}\right) \left(\frac{1}{\sqrt{d}}\sum_{y \in [d]} \bra{y,y}\right)$.
\end{itemize}

\begin{lemma}[\cite{ma2025construct}]
\label{lemma:Haar-twirling-1}
    $\begin{aligned}
        \E_{C \gets \mu^\Haar_d} \left[C^{\dag,\ot 2} \;\Pi^\eq\; C^{\ot 2}\right] = \frac{2}{d+1} \Pi^{\sym}.
    \end{aligned}$
\end{lemma}

\begin{dfn}[Partial transpose]
\label{dfn:partial-transpose}
    Let $O = \sum_{x,y,z,w} a_{x,y,z,w} \ket{x}\bra{y} \ot \ket{z}\bra{w}$ be an operator on $(\bbC^d)^{\ot 2}$. The partial transpose of $O$ is defined as
    $O^{T_2} := \sum_{x,y,z,w} a_{x,y,z,w} \ket{x}\bra{y} \ot \ket{w}\bra{z}.$
\end{dfn}

\begin{lemma}[\cite{ma2025construct}]
\label{lemma:Haar-twirling-2}
    $\begin{aligned}
    \left(\E_{C \gets \mu^\Haar_d} \left[C^{\dag,\ot 2} \;\Pi^\eq\; C^{\ot 2}\right]\right)^{T_2} - \Pi^{\EPR} = \frac{1}{d+1} (\bbmI - \Pi^\EPR).
    \end{aligned}$
\end{lemma}

\section{Subspace-Preserving Strong Pseudorandom Unitary}
\label{Sec:PRU}
Recently, Ma and Huang \cite{ma2025construct} constructed strong pseudorandom unitaries, which is a family of efficient $n$-qubit unitaries that are indistinguishable from $n$-qubit Haar random unitaries under forward and inverse queries.

\begin{theorem}[\cite{ma2025construct}]\label{thm:PRU-exists}
     Strong PRU exists assuming post-quantum one-way functions exist.
\end{theorem}

For our purposes, we need unitaries with both pseudorandomness and some other structure. In particular, we put forward the notion of \emph{subspace-preserving} pseudorandom unitaries, which are efficient unitaries that preserve every element in a linear subspace $S$ and behave pseudorandomly on the orthogonal space $S^\perp$. We recall the definition of $\mu^\Haar_{S^\perp}$ in \cref{def:haar}, which is a Haar random distribution over all unitary operators that preserve elements in $S$.
\begin{dfn}[Subspace-preserving strong pseudorandom unitary]
\label{def:spsPRU}
    An $S$-preserving strong pseudorandom unitary is a family $\set{\spPRU_k}_{k \in \cK}$ of $n$-qubit unitaries satisfying:
    \begin{itemize}
        \item Efficiency: There exists a $\poly(n)$-time quantum circuit that, when given any $k \in \cK$, implements $\spPRU_k$ up to an inverse-exponential error in diamond distance.
        \item $S$-preserving: For every $k \in \cK$ and $\ket{v} \in S$, it holds that $\spPRU_k \ket{v} = \ket{v}$.
        \item Pseudorandomness: For every $\poly(n)$-time binary-output adversary $\adv$ with queries in any order, we have
        \begin{align*}
            \left| \Pr_{k \leftarrow \cK}  \left[ 1 \gets \adv^{\spPRU_k,\; \spPRU_k^\dag} \right] - \Pr_{U \gets \mu^\Haar_{S^\perp}}  \left[ 1 \gets \adv^{U, U^\dag} \right] \right| \leq \negl(n).
        \end{align*}
    \end{itemize}
\end{dfn}

We only focus on subspaces of the form $S = \Span \set{\ket{x}}_{x \in \zo^n \setminus [d]}$ and always consider $d = n^{\omega(1)}$. Note that the strong PRU considered in \cite{ma2025construct} is equivalent to the special case of $S$-preserving strong pseudorandom unitary where $S = \Span \set{\ket{x}}_{x \in \zo^n \setminus [2^n]} = \Span \emptyset = \set{0}$.

Below, we review an intermediate theorem of \cite{ma2025construct} in \cref{sec:intermediate-PRU-thm}, and we show in \cref{sec:our-intermediate-PRU-thm} that a building block of it can be relaxed. We then construct $S$-preserving strong PRUs in \cref{sec:efficient-spsPRU}. In \cref{sec:path-recording-oracle}, we show how to efficiently simulate oracle access to a random unitary in $\mu^\Haar_{S^\perp}$.

\subsection{An Intermediate Theorem from \cite{ma2025construct}}
\label{sec:intermediate-PRU-thm}
We first recall an intermediate result by \cite{ma2025construct}. Note that the definition, theorem, and proof of this part in \cite{ma2025construct} directly extend to any finite dimension $d$ and need not be restricted to power-of-two dimensions. The restriction on the dimension was only required in \cite{ma2025construct} for an efficient instantiation of the $\PRU(\mathfrak{D})$ distribution.

\begin{dfn}[\cite{ma2025construct}, $\PRU(\mathfrak{D})$ distribution]
\label{dfn:PRU-distribution}
    For any distribution $\mathfrak{D}$ over $\calU(d)$, we define the distribution $\PRU({\frakD})$ as follows:
    \begin{enumerate}
        \item Sample a uniformly random permutation $\pi \gets \sSym_{d}$, a uniformly random $f \gets \{0,1, 2\}^d$, and two independently sampled unitaries $C, D \gets \mathfrak{D}$. Consider the unitaries 
        \begin{align*}
            F_f \coloneqq \sum_{x \in [d]} e^{2 \pi \cdot f(x) \cdot i/3} \ketbra{x} \quad \text{and} \quad P_{\pi} \coloneqq \sum_{x \in [d]} \ket{\pi(x)}\bra{x}.
        \end{align*}
        \item Output the unitary $\calO \coloneqq D \cdot P_\pi \cdot F_f \cdot C$.
    \end{enumerate} 
\end{dfn}

\begin{theorem}[\cite{ma2025construct}]
\label{thm:statistical-PRU-exact-2-design} Let $\frakD$ be an exact unitary $2$-design. Then for every $t$-query oracle adversary $\adv$ with queries in any order, we have
\begin{align*}
    \left| \Pr_{\calO \gets \PRU(\frakD)}  \left[ 1 \gets \adv^{\cO, \cO^\dag} \right] - \Pr_{\calO \gets \mu_d^{\Haar}} \left[ 1 \gets \adv^{\cO, \cO^\dag} \right] \right| \leq \frac{18 t(t+1)}{d^{1/8}}.
\end{align*}
\end{theorem}

The only place in \cite{ma2025construct} where the proof of \cref{thm:statistical-PRU-exact-2-design} depends on $\mathfrak{D}$ being an exact unitary $2$-design is through their Lemma 9.2. In their original proof of Lemma 9.2, the dependence on $\frakD$ being an exact unitary $2$-design appears in their proofs of Lemma 11.2 and Claim 28. Here, we isolate this dependence and restate it as a single lemma below.
Recall the definitions $\Pi^\eq \coloneqq \sum_{x \in [d]} \ketbra{x,x}$ and $\Pi^\EPR \coloneqq \left(\frac{1}{\sqrt{d}}\sum_{x \in [d]} \ket{x,x}\right) \left(\frac{1}{\sqrt{d}}\sum_{y \in [d]} \bra{y,y}\right)$.

\begin{lemma}[\cite{ma2025construct}, the proof and application of Lemma 9.2, restructured]
\label{lemma:ma2025-argument}${}$
\begin{itemize}
    \item For any exact unitary $2$-design $\frakD$ over $\cU(d)$ and integer $0 \leq t \leq d/2$, we have
    \begin{align}
        &\max_{\substack{\ell,r \geq 0:\\ \ell + r \leq t}}\; \frac{d}{d-\ell-r+1} \Biggl(\ell \norm{\E_{C \gets \frakD} C^{\dag,\ot 2} \;\Pi^\eq\; C^{\ot 2}}_\op \nonumber \\
        & \qquad\;\; + r \norm{\left(\E_{C \gets \frakD} (C^{\dag} \ot C^T) \;\Pi^\eq\; (C \ot \overline{C})\right) - \Pi^{\EPR}}_\op + 2r \sqrt{\frac{\ell + r}{d}} \Biggl) \le 6t \sqrt{\frac{t}{d}} \label{eq:pru-key-lemma-1}\\
        &\max_{\substack{\ell,r \geq 0:\\ \ell + r \leq t}} \; \frac{d}{d-\ell-r+1} \Biggl(\ell \norm{\E_{C \gets \frakD} C^{\ot 2} \;\Pi^\eq\; C^{\dag,\ot 2}}_\op \nonumber \\
        & \qquad\;\; + r \norm{\left(\E_{C \gets \frakD} (C \ot \overline{C}) \;\Pi^\eq\; (C^{\dag} \ot C^T)\right) - \Pi^{\EPR}}_\op + 2r \sqrt{\frac{\ell + r}{d}} \Biggl) \le 6t \sqrt{\frac{t}{d}} \label{eq:pru-key-lemma-2}
    \end{align}
    \item Let $\frakD$ be any distribution over $\cU(d)$ that satisfies (\ref{eq:pru-key-lemma-1}) and (\ref{eq:pru-key-lemma-2}). Then for every $t$-query oracle adversary $\adv$ with queries in any order, we have
\begin{align*}
    \left| \Pr_{\calO \gets \PRU(\frakD)}  \left[ 1 \gets \adv^{\cO, \cO^\dag} \right] - \Pr_{\calO \gets \mu_d^{\Haar}} \left[ 1 \gets \adv^{\cO, \cO^\dag} \right] \right| \leq \frac{18 t(t+1)}{d^{1/8}}.
\end{align*}
\end{itemize}
\end{lemma}

\subsection{Our Generalized Intermediate Theorem}
\label{sec:our-intermediate-PRU-thm}

We show a relaxed condition on $\frakD$ that is sufficient for $\PRU(\frakD)$ to be a statistical strong PRU. In particular, we only need the requirement that twirling a \emph{specific operator} by two copies over $\frakD$ is \emph{close to} twirling it by two copies over the Haar random distribution.

\begin{dfn}[Approximate restricted unitary $2$-design]

Let $O$ be an operator on $(\bbC^{d})^{\ot 2}$. 
A distribution $\frakD$ over $\cU(d)$ is called an $\eps$-approximate $O$-restricted unitary $2$-design if both of the following relations hold.
    \begin{equation}
    \norm{\E_{C \sim \frakD} \left[ C^{\ot 2}O C^{\dag,\ot 2} \right] - \E_{C \sim \mu^\Haar_d} \left[ C^{\ot 2}O C^{\dag,\ot 2} \right] }_{\op} \le \eps. 
    \label{eq:restricted-desing-requirement-1}
    \end{equation}
    \begin{equation}
    \norm{\E_{C \sim \frakD} \left[ C^{\dag,\ot 2}O C^{\ot 2} \right] - \E_{C \sim \mu^\Haar_d} \left[ C^{\dag,\ot 2}O C^{\ot 2} \right] }_{\op} \le \eps.
    \label{eq:restricted-desing-requirement-2}
    \end{equation}
\end{dfn}
In our theorem, the specific operator of interest is $\Pi^\eq := \sum_{x\in [d]} \ketbra{x,x}$.

\begin{thm}
\label{thm:our-statistical-PRU} Let $\frakD$ be a $\frac{1}{d(d+1)}$-approximate $\Pi^\eq$-restricted unitary $2$-design. Then for every $t$-query oracle adversary $\adv$ with queries in any order, we have
\begin{align*}
    \left| \Pr_{\calO \gets \PRU(\frakD)}  \left[ 1 \gets \adv^{\cO, \cO^\dag} \right] - \Pr_{\calO \gets \mu_d^{\Haar}} \left[ 1 \gets \adv^{\cO, \cO^\dag} \right] \right| \leq \frac{18 t(t+1)}{d^{1/8}}.
\end{align*}
\end{thm}

Before proving \cref{thm:our-statistical-PRU}, we show a lemma for partial transpose (\cref{dfn:partial-transpose}).
\begin{lemma}\label{lemma:operator-norm-after-partial-transpose}
    For any operator $O$ on $(\bbC^d)^{\ot 2}$, we have
    $\norm{O^{T_2}}_\op \le d\cdot \norm{O}_\op$.
\end{lemma}
\begin{proof}
    For any unit vectors $\ket{u}, \ket{v} \in (\bbC^{d})^{\ot 2}$, we can expand them as
    $\ket{u} = \sum_{x \in [d]} \ket{u_x} \ket{x}$ and $\ket{v} = \sum_{y \in [d]} \ket{v_y} \ket{y}$. By the cyclicity property of trace, the fact that partial transpose preserves trace, and the H\"older inequality, we have
    \begin{align*}
        \abs{\bra{u} O^{T_2} \ket{v}} = \abs{\Tr(O^{T_2} \ket{v}\bra{u})} = \abs{\Tr(O (\ket{v}\bra{u})^{T_2})} \le \norm{O}_\op \norm{(\ket{v}\bra{u})^{T_2}}_1
    \end{align*}
    Note that we have
    \begin{align*}
        \mathsf{rank}\left((\ket{v}\bra{u})^{T_2}\right) = \mathsf{rank}\left(\sum_{x,y\in[d]} \ket{v_y}\bra{u_x} \ot \ket{x}\bra{y}\right) \le d^2
    \end{align*}
    By the Cauchy-Schwarz inequality, we have
    \begin{align*}
        \norm{(\ket{v}\bra{u})^{T_2}}_1 \le \sqrt{\mathsf{rank}\left((\ket{v}\bra{u})^{T_2}\right)} \norm{(\ket{v}\bra{u})^{T_2}}_2 \le d,
    \end{align*}
    where we use the fact that the squared $2$-norm equals the sum of the squared absolute values of all coefficients, and is therefore invariant under partial transpose. Hence, we get
    \begin{align*}
        \norm{O^{T_2}}_\op = \max_{\text{unit }\ket{u},\ket{v} \in (\bbC^{d})^{\ot 2}} \abs{\bra{u} O^{T_2} \ket{v}}  \le \norm{O}_\op \max_{\text{unit }\ket{u},\ket{v} \in (\bbC^{d})^{\ot 2}} \norm{(\ket{v}\bra{u})^{T_2}}_1 \le d \norm{O}_\op
    \end{align*}
\end{proof}

\begin{proof}[Proof of \cref{thm:our-statistical-PRU}]
    We prove that $\frakD$ also satisfies the condition (\ref{eq:pru-key-lemma-1}) stated in \cref{lemma:ma2025-argument}.
    Since $\frakD$ is a $\frac{1}{d(d+1)}$-approximate $\Pi^\eq$-restricted $2$-design, we have
    $$\norm{\E_{C \sim \frakD} \left[ C^{\dag,\ot 2} \;\Pi^\eq\; C^{\ot 2} \right] - \E_{C \sim \mu^\Haar_d} \left[ C^{\dag,\ot 2} \;\Pi^\eq\; C^{\ot 2} \right] }_{\op} \le \frac{1}{d(d+1)}
    $$
    By \cref{lemma:operator-norm-after-partial-transpose}, we also have
    \begin{align*}
        &\norm{\left(\left(\E_{C \sim \frakD} \left[ C^{\dag,\ot 2} \;\Pi^\eq\; C^{\ot 2} \right]\right)^{T_2} - \Pi^\EPR\right) - \left(\left(\E_{C \sim \mu^\Haar_d} \left[ C^{\dag,\ot 2} \;\Pi^\eq\; C^{\ot 2} \right]\right)^{T_2} - \Pi^\EPR\right) }_{\op}\\
        =& \norm{\left(\E_{C \sim \frakD} \left[ C^{\dag,\ot 2} \;\Pi^\eq\; C^{\ot 2} \right]\right)^{T_2} - \left(\E_{C \sim \mu^\Haar_d} \left[ C^{\dag,\ot 2} \;\Pi^\eq\; C^{\ot 2} \right]\right)^{T_2} }_{\op}\\
        \le& \;d \cdot \norm{\E_{C \sim \frakD} \left[ C^{\dag,\ot 2} \;\Pi^\eq\; C^{\ot 2} \right] - \E_{C \sim \mu^\Haar_d} \left[ C^{\dag,\ot 2} \;\Pi^\eq\; C^{\ot 2} \right] }_{\op} \le \frac{1}{d+1}
    \end{align*}
    Therefore, we have
    \begin{align*}
        &\ell \norm{\E_{C \gets \frakD} C^{\dag,\ot 2} \;\Pi^\eq\; C^{\ot 2}}_\op + r \norm{\left(\E_{C \gets \frakD} (C^{\dag} \ot C^T) \;\Pi^\eq\; (C \ot \overline{C})\right) - \Pi^{\EPR}}_\op\\
        =&\ell \norm{\E_{C \gets \frakD} C^{\dag,\ot 2} \;\Pi^\eq\; C^{\ot 2}}_\op + r \norm{\left(\E_{C \gets \frakD} (C^{\dag,\ot 2}) \;\Pi^\eq\; C^{\ot 2}\right)^{T_2} - \Pi^{\EPR}}_\op\\
        \le& \;\ell \left(\norm{\E_{C \gets \mu^\Haar_d} C^{\dag,\ot 2} \;\Pi^\eq\; C^{\ot 2}}_\op + \frac{1}{d(d+1)}\right) \\
        &\quad + r \left(\norm{\left(\E_{C \gets \mu^\Haar_d} C^{\dag,\ot 2} \;\Pi^\eq\; C^{\ot 2}\right)^{T_2} - \Pi^{\EPR}}_\op + \frac{1}{d+1}\right)\\
        =&\; \ell \left(\norm{\frac{2}{d+1} \Pi^{\sym}}_\op + \frac{1}{d(d+1)}\right)  + r \left(\norm{\frac{1}{d+1} (\bbmI - \Pi^\EPR)}_\op + \frac{1}{d+1}\right) 
        \le \frac{3 \ell + 2 r}{d+1}
    \end{align*}
    where the last equality uses \cref{lemma:Haar-twirling-1,lemma:Haar-twirling-2} for the result of taking the Haar average.
    Hence, for our given $\frakD$, the left-hand side term of inequality (\ref{eq:pru-key-lemma-1}) is upper bounded by
    \begin{align*}
        & \max_{\substack{\ell,r \geq 0:\\ \ell + r \leq t}}\;  \frac{d}{d-\ell-r+1} \left(\frac{3\ell + 2r}{d+1} + 2r \sqrt{\frac{\ell + r}{d}} \right)\\
        \le& \frac{d}{d-t+1}\left(\frac{2t}{d+1} + 2 t \sqrt{\frac{t}{d}} \right) \le \frac{d}{d-t} \cdot 3t \sqrt{\frac{t}{d}} \le 6t \sqrt{\frac{t}{d}}
    \end{align*}
    The first inequality holds by the observation that the maximum is achieved at $\ell =0, r=t$. The next inequality holds because $\frac{2}{d+1} \le \sqrt{\frac{1}{d}}$ for every $d$, and the last inequality follows from $t \le \frac{d}{2}$. We have therefore proved that (\ref{eq:pru-key-lemma-1}) holds for $\frakD$, and (\ref{eq:pru-key-lemma-2}) holds for $\frakD$ by a similar argument. Then we conclude our proof by applying the second item of \cref{lemma:ma2025-argument}.
\end{proof}

\subsection{Efficient Construction}
\label{sec:efficient-spsPRU}

To build an $S$-preserving strong pseudorandom unitary for $S = \Span \set{\ket{x}}_{x \in \zo^n \setminus [d]}$, we seek an efficient family of $n$-qubit unitaries that acts as the identity on $S$ and has behavior computationally indistinguishable from the $\PRU(\frakD)$ distribution (\cref{dfn:PRU-distribution}) on $S^\perp = \Span\set{\ket{x}}_{x \in [d]}$ with a suitable choice of the distribution $\frakD$. Recall that $\PRU(\frakD)$ samples a random function $f:[d] \to \set{0,1,2}$, a random permutation $\pi:[d] \to [d]$, and two unitary operators $C,D$ from $\frakD$, and outputs the unitary $\cO = D \cdot P_\pi \cdot F_f \cdot C$.

Efficiently simulating a family of permutations that preserve elements in $S$ and are pseudorandom on $[d]$ (instead of $\zo^n$) requires some care. In \cref{sec:prp}, we show how to construct such a family with post-quantum security. In \cref{sec:design}, we construct an efficient distribution $\frakD$ of $n$-qubit unitaries that preserves elements in $S$ and is an approximate $\Pi^\eq$-restricted $2$-design on $S^\perp = \Span\set{\ket{x}}_{x \in [d]}$. We combine these components and build our $S$-preserving strong PRU in \cref{sec:spsPRU-final-construction}.

\subsubsection{Post-Quantum Pseudorandom Permutation Over $[d]$}
\label{sec:prp}

A post-quantum pseudorandom permutation (PRP) is an efficiently computable keyed permutation that is indistinguishable from a truly random permutation to any efficient adversary with oracle access to the permutation and its inverse. Moreover, given the key, the permutation can be efficiently inverted. Here, we focus on PRPs that permute a set $[d]$ which is not necessarily the set of all $n$-bit strings. In the literature, this concept is often referred to as format-preserving encryption (FPE) \cite{black2002ciphers}.

\begin{dfn}[Post-quantum pseudorandom permutation (PRP)]
\label{dfn:prp}
Let $d = d(\secparam)$. 
A post-quantum pseudorandom permutation over $[d]$ is a pair of efficient deterministic classical algorithms
$\PRP, \PRP^{-1} : \{0,1\}^\lambda \times [d] \to [d]$ such that:

\begin{itemize}
    \item $\PRP(k,\cdot), \PRP^{-1}(k,\cdot)$ are permutations over $[d]$ for every $k \in \zo^\secparam$.
    \item $\PRP(k,\cdot),\PRP_k^{-1}(k,\cdot)$ are inverses of each other for every $k \in \zo^\secparam$.
    \item For every quantum polynomial-time adversary $\adv$ that makes quantum queries to its given oracles, we have
	\[\left|\Pr_{k \gets \zo^\secparam}\left[\adv^{\PRP(k,\cdot),\PRP^{-1}(k,\cdot)}(1^\secparam) =1\right]-\Pr\left[\adv^{P,P^{-1}}(1^\secparam)=1\right]\right|<\negl(\secparam),\]
    where $P$ is a random permutation over $[d]$. 
\end{itemize}
\end{dfn}

The work of \cite{morris2014sometimes} showed how to construct a statistically secure pseudorandom permutation over any prescribed set $[d]$ from a random function, where the security holds even against an unbounded adversary that can query the entire permutation table. This result was later reformulated as the construction of a full-domain function-to-permutation converter \cite{zhandry2025note} and applied to build post-quantum PRPs over domains of size $2^n$. Our interest lies in post-quantum PRPs over arbitrary domain sizes $d$, which we obtain based on \cite{morris2014sometimes} and \cite{zhandry2025note}. In what follows, we adapt the definition from \cite{zhandry2025note} to more precisely capture that the result of \cite{morris2014sometimes} holds for arbitrary domain sizes $d$.

\begin{dfn}
\label{dfn:full-domain-converter}
Let $d = d(\secparam)$. A full-domain function-to-permutation converter over $[d]$ is a pair $(\FPC,\FPCInv)$ of efficient deterministic classical oracle algorithms where:

\begin{itemize}
    \item $\FPC,\FPCInv$ take as input the security parameter $1^\lambda$ and an integer $x\in [d]$, run in $\poly(\secparam)$ time with $\poly(\secparam)$ classical queries to a function $O:\{0,1\}^{m(\secparam)}\rightarrow\{0,1\}^{m'(\secparam)}$ where $m(\secparam),m'(\secparam) = \poly(\secparam)$,
    and output an integer $y\in [d]$.
	\item $\FPC,\FPCInv$ are inverses: $\FPCInv^O(1^\lambda,\FPC^O(1^\lambda,x))=x$ for all $x\in[d]$ and all functions $O$.
	\item $\FPC,\FPCInv$ are indistinguishable from a random permutation and its inverse under arbitrary oracle access. That is, for any (possibly inefficient) adversary $\adv$ that makes arbitrary queries to its given oracles, we have
	\[\left|\Pr\left[\adv^{\FPC^O(1^\lambda,\cdot),\FPCInv^O(1^\lambda,\cdot)}(1^\lambda)=1\right]-\Pr\left[\adv^{P,P^{-1}}(1^\lambda)=1\right]\right|<\negl(\secparam),\]
    where $O$ is a random function, and $P$ is a random permutation over $[d]$. 
\end{itemize}
\end{dfn}

\begin{lemma}[\cite{morris2014sometimes}]\label{lem:fulldomain} Full-domain function-to-permutation converter over $[d]$ exists for every (not necessarily a power of two) $d = d(\secparam) \le O(2^{\poly(\secparam)})$.
\end{lemma}

By combining full-domain function-to-permutation converters over $[d]$ with post-quantum PRFs, one directly obtains post-quantum PRPs over $[d]$.

\begin{lemma}
    If post-quantum PRFs exist, then post-quantum PRP over $[d]$ exists for every (not necessarily a power of two) $d = d(\secparam) \le O(2^{\poly(\secparam)})$.
\end{lemma}

\begin{proof}
Let $(\FPC, \FPCInv)$ be a full-domain function-to-permutation converter over $[d]$ that queries functions of the form  $\zo^{m(\secparam)} \to \zo^{m'(\secparam)}$. Let $\PRF: \zo^{\secparam} \times \zo^{m(\secparam)} \to \zo^{m'(\secparam)}$ be a post-quantum pseudorandom function. 
One can construct a post-quantum pseudorandom permutation $\PRP,\PRP^{-1}: \zo^\secparam \times [d] \to [d]$ as follows. $$\PRP(k,x) := \FPC^{\PRF(k,\cdot)}(1^\secparam,x) \text{ and } \PRP^{-1}(k,x) := \FPCInv^{\PRF(k,\cdot)}(1^\secparam,x).$$
It is clear that $\PRP(k,\cdot), \PRP^{-1}(k,\cdot)$ are permutations over $[d]$ that are inverses of each other for every $k$ by the definition of $(\FPC,\FPCInv)$.
To prove security, let $\adv$ be any quantum polynomial-time adversary and consider the following quantities, where we sample a random key $k \in \zo^\secparam$, a random function $O: \zo^m \to \zo^{m'}$, and a random permutation $P: [d] \to [d]$.
\begin{align*}
    G_0(\secparam) &:= \Pr_k\left[\adv^{\PRP(k,\cdot),\PRP^{-1}(k,\cdot)}(1^\lambda)=1\right] = \Pr_k\left[\adv^{\FPC^{\PRF(k,\cdot)}(1^\lambda,\cdot),\FPCInv^{\PRF(k,\cdot)}(1^\lambda,\cdot)}(1^\lambda)=1\right]\\
    G_1(\secparam) &:= \Pr_O\left[\adv^{\FPC^{O(\cdot)}(1^\lambda,\cdot),\FPCInv^{O(\cdot)}(1^\lambda,\cdot)}(1^\lambda)=1\right]\\
    G_2(\secparam) &:=\Pr_P\left[\adv^{P,P^{-1}}(1^\lambda)=1\right]
\end{align*}
We have $\abs{G_0(\secparam) - G_1(\secparam)} = \negl(\secparam)$ by the post-quantum security of $\PRF$ and the fact that $\adv, \FPC, \FPCInv$ run in polynomial time. We also have $\abs{G_1(\secparam) - G_2(\secparam)} = \negl(\secparam)$ by the security of $(\FPC,\FPCInv)$. Hence, $\abs{G_0(\secparam) - G_2(\secparam)} = \negl(\secparam)$, which proves that $\PRP$ is a post-quantum PRP.
\end{proof}

\subsubsection{Subspace-Preserving Approximate Restricted $2$-Design}
\label{sec:design}

Let $S$ be a subspace in $\bbC^{2^n}$ and $O$ be an operator on $S^\perp$.
\begin{dfn}
An $S$-preserving $\eps$-approximate $O$-restricted unitary $2$-design on $n$ qubits is a distribution $\frakD$ over $n$-qubit unitary operators such that
\begin{itemize}
    \item Every $U \in \Supp(\frakD)$ can be decomposed as $U = \identity_S \oplus U_{S^\perp}$ where $U_{S^\perp} \in \cU(S^\perp)$. In particular, we have $U \ket{s} = \ket{s}$ for every $U \in \Supp(\frakD)$ and unit vector $\ket{s} \in S$.
    \item The distribution $\frakD|_{S^\perp}$ given by $U_{S^\perp}$ where $U \gets \frakD$ is an $\eps$-approximate $O$-restricted unitary $2$-design over $\cU(S^\perp)$.
\end{itemize}
\end{dfn}

Clearly, any $\eps$-approximate unitary $2$-design on $\cU(2^n)$ automatically satisfies the above definition by setting $S$ to be the trivial subspace and $O$ to be any operator. For our purposes, we need an ensemble that preserves some given non-trivial subspace $S$ and approximates the Haar distribution over $\cU(S^\perp)$ on a specific operator $O$.

Let $d \le 2^n$ and $\omega_d = e^{2\pi i / d}$.
In the rest of this subsection, we only consider the subspace $S = \Span \set{\ket{x}}_{x \in \zo^n \setminus [d]}$ and the operator $O = \Pi^\eq = \sum_{x \in [d]} \ketbra{x,x}$. We will obtain an $S$-preserving approximate $\Pi^\eq$-restricted $2$-design by adapting the construction of \cite{nakata2017unitary}.

\begin{dfn}\label{dfn:three-fold-ensemble}
    Define $\threefold$ as the $n$-qubit unitary ensemble $\set{Z_{f_3} H_d Z_{f_2} H_d Z_{f_1}}$ for uniformly and independently chosen random functions $f_1,f_2,f_3: [d] \to [d]$, where the $n$-qubit phase operator $Z_f$ associated with a function $f: [d] \to [d]$ is defined as
$$Z_f : \ket{x} \mapsto \begin{cases}
    \omega_d^{f(x)} \ket{x} & \text{if } x \in [d] \\
    \ket{x} & \text{otherwise}
\end{cases},$$
and the $n$-qubit Fourier transform $H_d$ is defined as
$$H_d : \ket{x} \mapsto \begin{cases}
    \frac{1}{\sqrt{d}} \sum_{y \in [d]} \omega_d^{xy} \ket{y} & \text{if } x \in [d] \\
    \ket{x} & \text{otherwise}
\end{cases}.$$
\end{dfn}
Recall that $S = \Span \set{\ket{x}}_{x \in \zo^n \setminus [d]}$ and $\Pi^\eq = \sum_{x \in [d]} \ketbra{x,x}$.
\begin{lemma}
\label{lemma:approximate-restricted-design}
    $\threefold$ is an $S$-preserving $\frac{1}{d(d+1)}$-approximate $\Pi^\eq$-restricted $2$-design on $n$ qubits.
\end{lemma}

To prove the lemma, we also define the ensemble ${\frakD}_{\mathsf{std}}$ as $\set{Z_f}$ for a uniform $f:[d] \to [d]$. The projection onto the symmetric subspace can be expressed as $\Pi^{\sym} = \frac{1}{2}\left(\bbmI + \bbmF\right) = \Pi^\eq + \Pi'$, where $\Pi' = \sum_{x>y} \frac{\ket{x,y} + \ket{y,x}}{\sqrt{2}} \frac{\bra{x,y} + \bra{y,x}}{\sqrt{2}}$ is a projector orthogonal to $\Pi^\eq$.


\begin{proof}[Proof of \cref{lemma:approximate-restricted-design}]
    By construction, both $Z_f$ and $H_d$ preserve all vectors in $S$. Therefore, every unitary in the support of $\threefold$ also preserves all vectors in $S$. Let us gradually compute the $2$-wise twirling $\fulltwirl(\Pi^\eq) = \xtwirl\xtwirl\ztwirl(\Pi^\eq)$.
    By definition,
\[
\ztwirl(\ket{x,y}\bra{x',y'}) = \E_{f}\left[Z_{f}^{\ot 2} \ket{x,y}\bra{x',y'} Z_{f}^{\dagger, \ot 2}\right] = \mathbb{E}_f\left[\omega_d^{f(x)+f(y)-f(x')-f(y')}\right] \ket{x,y}\bra{x',y'}
\]
The last expectation is 1 if and only if $\{x,y\} = \{x',y'\}$, and 0 otherwise, so we have
\begin{align*}
        \ztwirl(\ket{x,y}\bra{x',y'}) = \begin{cases}
        \ket{x,y}\bra{x,y} & \text{if } (x,y)=(x',y')\\
        \ket{x,y}\bra{y,x} & \text{if } (x,y)=(y',x')\\
        0 & \text{otherwise}
        \end{cases}
    \end{align*}
Hence $\ztwirl(\Pi^\eq) = \Pi^\eq$. Next, by expanding the Fourier transformation $\Hg_d$ and using the above expression for $\ztwirl(\cdot)$, we get
\begin{align*}
\xtwirl(\ket{x,y}\bra{y,x}) 
&= \ztwirl \left(H_d^{\ot 2} \ket{x,y}\bra{y,x} H_d^{\dagger, \ot 2}\right)\\
&= \ztwirl \left(\frac{1}{d^2} \sum_{a,b,a',b'} \omega_d^{x(a-b') + y(b-a')} \ket{a,b}\bra{a',b'} \right)\\
&= \frac{1}{d^2} \left(\sum_{a,b} \ket{a,b}\bra{b,a} + \sum_{a \neq b} \omega_d^{x(a-b') + y(b-a')} \ket{a,b}\bra{a,b}\right)\\
&= \frac{1}{d^2} \left(\bbmF +  \sum_{a \neq b} \omega_d^{(x-y)(a-b)}  \ket{a,b}\bra{a,b}\right)
\end{align*}
Summing over all $x,y \in [d]$ gives $\xtwirl(\bbmF) = \bbmF$. Summing over $x = y \in [d]$ gives
\begin{align*}
    \xtwirl(\Pi^\eq)&= \sum _{x \in [d]}\xtwirl(\ketbra{x,x})\\
    &= d \cdot \frac{1}{d^2} \left(\bbmF + \sum_{a \neq b} \ket{a,b}\bra{a,b}\right) \\
    &= \frac{1}{d} \left(\bbmF + \sum_{a,b} \ket{a,b}\bra{a,b} - \sum_a \ketbra{a,a}\right) 
    = \frac{1}{d} \left(\bbmI + \bbmF - \Pi^\eq \right)
\end{align*}
Therefore, we have
    \begin{align*}
        \fulltwirl(\Pi^\eq) &= \xtwirl\xtwirl\ztwirl\left(\Pi^\eq\right) \\
        &= \xtwirl\left(\frac{1}{d}\left(\bbmI + \bbmF - \Pi^\eq \right)\right)\\
        &= \frac{1}{d}\left(\bbmI + \bbmF - \frac{1}{d}\left(\bbmI + \bbmF - \Pi^\eq \right) \right)\\
        &= \frac{d-1}{d^2} \cdot 2(\Pi^\eq + \Pi') + \frac{1}{d^2} \Pi^\eq \\
        &= \frac{2d-1}{d^2} \Pi^\eq + \frac{2d-2}{d^2} \Pi'
    \end{align*}
    By \cref{lemma:Haar-twirling-1}, we have $\twirl^{(2)}_{\mu^\Haar_{S^\perp}}(\Pi^\eq) = \frac{2}{d+1} \Pi^\sym = \frac{2}{d+1} \Pi^\eq + \frac{2}{d+1} \Pi'$. Therefore, we have
    \begin{align*}
        \norm{\fulltwirl\left(\Pi^\eq\right) - \twirl^{(2)}_{\mu^\Haar_{S^\perp}}(\Pi^\eq)}_\op 
        =& \norm{\left(\frac{2d-1}{d^2} - \frac{2}{d+1}\right) \Pi^\eq + \left(\frac{2d-2}{d^2} - \frac{2}{d+1}\right)\Pi'}_\op \nonumber\\
        =& \norm{\frac{d-1}{d^2(d+1)} \Pi^\eq - \frac{2}{d^2(d+1)}\Pi'}_\op
        \le \frac{1}{d(d+1)}
    \end{align*}
This shows that $\threefold|_{S^\perp}$ satisfies the first requirement (\ref{eq:restricted-desing-requirement-1}) for being a $\frac{1}{d(d+1)}$-approximate $\Pi^\eq$-restricted $2$-design. To show the second requirement (\ref{eq:restricted-desing-requirement-2}), note that the inverse of a unitary sampled from $\threefold$ is distributed as the unitary ensemble $\set{Z_{f_3} H_d^\dag Z_{f_2} H_d^\dag Z_{f_1}}$ for random functions $f_1,f_2,f_3$, and that the inverse of a Haar random unitary is a Haar random unitary. The same derivation as above shows that the latter ensemble satisfies (\ref{eq:restricted-desing-requirement-1}), which is equivalent to the statement that $\threefold|_{S^\perp}$ satisfies (\ref{eq:restricted-desing-requirement-2}). Hence, we have shown that $\threefold$ is an $S$-preserving $\frac{1}{d(d+1)}$-approximate $\Pi^\eq$-restricted $2$-design.
\end{proof}

\subsubsection{Subspace-Preserving Strong Pseudorandom Unitary}
\label{sec:spsPRU-final-construction}


\begin{theorem}
\label{thm:spsPRU}
There exists an $S$-preserving strong pseudorandom unitary (\cref{def:spsPRU}) for $S = \Span\set{\ket{x}}_{x \in \zo^n \setminus [d]}$ and any $d = n^{\omega(1)}$, assuming the existence of post-quantum one-way functions.
\end{theorem}

\begin{Construction}{$\Span\set{\ket{x}}_{x \in \zo^n \setminus [d]}$-preserving strong pseudorandom unitary.}{}\\
\label{construction:spPRU}
{\bf Ingredients.}
\begin{itemize}
    \item Post-quantum pseudorandom permutation $\PRP: \zo^{n} \times [d] \to [d]$.
    \item Post-quantum pseudorandom function $\PRF: \zo^n \times [d] \to \set{0,1,2}$.
    \item Post-quantum pseudorandom function $\PRF':\zo^n \times \Big(\zo \times \set{1,2,3} \times [d]\Big) \to [d]$.
\end{itemize}

{\bf Unitary $\spPRU_{k \in \zo^{3n}}$.}
    \begin{itemize}
        \item Parse $k = (k_{\PRP}, k_{\PRF}, k_{\PRF'})$.
        \item Set $\pi(x) = \begin{cases}
            \PRP(k_{\PRP},x) & \text{ if } x \in [d]\\
            x & \text{ otherwise}
        \end{cases} \text{ and } f(x) = \begin{cases}
            \PRF(k_{\PRF},x) & \text{ if } x \in [d]\\
            0 & \text{ otherwise}
        \end{cases}.$
        \item Set $f'_{i,j}(x) = \PRF'(k_{\PRF'}, (i,j,x))$.
        \item Compute $\widehat{C} = Z_{f'_{0,3}} \cdot H_d \cdot Z_{f'_{0,2}} \cdot H_d \cdot Z_{f'_{0,1}}$ and $\widehat{D} = Z_{f'_{1,3}} \cdot H_d \cdot Z_{f'_{1,2}} \cdot H_d \cdot Z_{f'_{1,1}}$.
        \item Output $\spPRU_k := \widehat{D} \cdot P_\pi \cdot F_f \cdot \widehat{C}$. 
    \end{itemize}
\end{Construction}

\begin{proof}
    We prove that Construction \ref{construction:spPRU} is a subspace-preserving strong PRU. By the efficiency of the PRP and PRFs, we know that $\pi, \pi^{-1}, f, f'_{i,j}$ are efficiently computable, and thus $P_\pi, F_f, Z_{f'_{i,j}}$ are efficiently computable. The Fourier transform $H_d$ can be efficiently computed within any inverse-exponential error \cite{kitaev1995quantum,hales2000improved}. Since $\spPRU_k$ is defined as the composition of these unitaries, it is also $\poly(n)$-time computable within an inverse-exponential error. $\mathsf{spsPRU}_k$ is $S$-preserving because its components $P_\pi, F_f, Z_{f'_{i,j}}, H_d$ are all $S$-preserving by construction. To show that $\spPRU$ is indistinguishable from $\mu^\Haar_{S^{\perp}}$, consider the following hybrids.
    \begin{itemize}
    \item $\hybrid_0$: Produce the unitary $\spPRU_k$ as in Construction \ref{construction:spPRU} with a uniformly random $k$.
    \item $\hybrid_1$: Replace $\PRF(k_\PRF,\cdot)$ in $\hybrid_0$ with a truly random function.
    \item $\hybrid_2$: Replace $\PRF'(k_{\PRF'},\cdot)$ in $\hybrid_1$ with a truly random function.
    \item $\hybrid_3$: Replace $\PRP(k_\PRP,\cdot)$ in $\hybrid_2$ with a truly random permutation in $\Sym_d$.
    \item $\hybrid_4$: Output a unitary sampled from $\mu^\Haar_{S^\perp}$.
    \end{itemize}
    By the post-quantum security of $\PRF, \PRF', \PRP$ and the fact that random functions can be perfectly and efficiently simulated for a bounded number of queries, any quantum polynomial-time adversary with forward and inverse queries to the given unitary has negligible advantage in distinguishing between hybrids $\hybrid_0, \hybrid_1, \hybrid_2, \hybrid_3$. The unitary produced by $\hybrid_3$ is $S$-preserving. Hence, its restriction to $S^\perp$ is still a unitary. By construction, the restricted unitary follows the distribution $\PRU(\threefold|_{S^\perp})$. By \cref{lemma:approximate-restricted-design}, $\threefold|_{S^\perp}$ is a $\frac{1}{d(d+1)}$-approximate $\Pi^{\eq}$-restricted $2$-design over $\cU(S^\perp)$. By \cref{thm:our-statistical-PRU} and the condition that $d = n^{\omega(1)}$, any quantum polynomial-time adversary with forward and inverse queries has negligible advantage in distinguishing between $\PRU(\threefold|_{S^\perp})$ and $\mu^\Haar_{d}$. Therefore, indistinguishability holds between $\hybrid_3, \hybrid_4$. Altogether, $\hybrid_0$ is indistinguishable from $\hybrid_4$, and thus the constructed $\spPRU$ is indistinguishable from $\mu^\Haar_{S^\perp}$ against any quantum polynomial-time adversary with forward and inverse queries. This completes the proof.
\end{proof}

\subsection{The Subspace-Preserving Path-Recording Oracle}
\label{sec:path-recording-oracle}
The path-recording oracle \cite{ma2025construct} provides an efficient way to simulate oracle access to Haar random unitaries. Here, we demonstrate that we can obtain an efficient simulation for $\mu^\Haar_{S^\perp}$ as well, where $S = \Span\set{\ket{x}}_{x \in \zo^n \setminus [d]}$.
We first recall some definitions and results from \cite{ma2025construct} that directly extend to any finite dimension $d$ since they make no reference to qubit representations.

For a multiset $R = \{(x_1,y_1),\dots,(x_t,y_t)\}$ of ordered pairs $(x_j,y_j) \in [d]^2$ of length $t$, we define $\Dom(R) = \{x \in [d] : \exists y \ \text{s.t.} (x,y) \in R\}$ and $\Im(R) = \{y \in [d] : \exists x \ \text{s.t.} (x,y) \in R\}$, and we define $\ket{R}$ as the normalization of $\sum_{\pi \in \sSym_t} \ket{x_{\pi(1)},y_{\pi(1)},\dots,x_{\pi(t)},y_{\pi(t)}}$. Consider two variable-length registers $\gsL,\gsR$ that each hold a state over the Hilbert space $\bigoplus_{t=0}^{d} (\bbC^d \ot \bbC^d)^{\ot t}$, where states of the form $\ket{x_1,y_1,\dots,x_t,y_t}$ span the subspace $(\bbC^d\ot \bbC^d)^{\ot t}$ for each $t$. By convention, the corresponding $1$-dimensional subspace for $t = 0$ is spanned by $\ket{\emptyset}$.
\begin{dfn}[\cite{ma2025construct}, Path-recording oracle]
\label{def:symmetric-V}
    Define the linear operators $\PRO^\mathrm{L}, \PRO^\mathrm{R}$ such that for every $x \in [d]$ and multisets $L$ and $R$ of length less than $d$,
    \begin{align*}
        \PRO^\mathrm{L} \cdot \ket{x}_{\gsA} \ket{L}_{\gsL} \ket{R}_{\gsR} &\coloneqq \sum_{\substack{y \in [d]:\\ y\not\in \Im(L \cup R)}} \frac{1}{\sqrt{d - \abs{\Im(L \cup R)}}} \ket{y}_{\gsA} \ket{L \cup \{(x,y)\}}_{\gsL} \ket{R}_{\gsR},\\
        \PRO^\mathrm{R} \cdot \ket{x}_{\gsA} \ket{L}_{\gsL} \ket{R}_{\gsR} &\coloneqq \sum_{\substack{y \in [d]:\\ y\not\in \Dom(L \cup R)}} \frac{1}{\sqrt{d - \abs{\Dom(L \cup R)}}} \ket{y}_{\gsA} \ket{L}_{\gsL} \ket{R \cup \{(y, x)\} }_{\gsR}.
    \end{align*}
    The path-recording oracle $\PRO$ is defined as the operator
    \begin{align*}
        \PRO &\coloneqq \PRO^\mathrm{L} \cdot (\Id - \PRO^\mathrm{R} \cdot \PRO^{\mathrm{R},\dagger}) + (\Id - \PRO^\mathrm{L} \cdot \PRO^{\mathrm{L},\dagger}) \cdot \PRO^{\mathrm{R},\dagger}.
    \end{align*}
\end{dfn}

\begin{lemma}[\cite{ma2025construct}]
\label{lemma:PRO-from-MH25} For every $t$-query oracle adversary $\adv$ with queries in any order,
\begin{align*}
    \left| \Pr_{\calO \gets \mu^\Haar_d}  \left[ 1 \gets \adv^{\cO, \cO^\dag} \right] - \Pr_{\gsL \gets \ket{\emptyset}, \gsR \gets \ket{\emptyset}} \left[ 1 \gets \adv^{\PRO, \PRO^\dag} \right] \right| \leq \frac{9 t(t+1)}{d^{1/8}}.
\end{align*}
\end{lemma}

We would like to simulate $\mu^{\Haar}_{S^{\perp}} = \mu^{\Haar}_{\Span\set{\ket{x}}_{x\in [d]}}$. To this end, we extend the definition of $\PRO^\mathrm{L}, \PRO^\mathrm{R}$ from \cref{def:symmetric-V} so that for every $x \in \zo^n \setminus [d]$ and multisets $L,R$,
\begin{align}
        \PRO^\mathrm{L} \cdot \ket{x}_{\gsA} \ket{L}_{\gsL} \ket{R}_{\gsR} &\coloneqq \ket{x}_{\gsA} \ket{L}_{\gsL} \ket{R}_{\gsR},
        \label{eq:extend-PRO-1}\\
        \PRO^\mathrm{R} \cdot \ket{x}_{\gsA} \ket{L}_{\gsL} \ket{R}_{\gsR} &\coloneqq \ket{x}_{\gsA} \ket{L}_{\gsL} \ket{R}_{\gsR}. \label{eq:extend-PRO-2}
    \end{align}
    We define $\PRO$ from $\PRO^\mathrm{L}, \PRO^\mathrm{R}$ in the same manner, and we call it the $S$-preserving path-recording oracle. Then $\PRO$ acts as the identity on $S = \Span\set{\ket{x}}_{x \in \zo^n \setminus [d]}$ and the restriction of $\PRO$ to $S^\perp$ is the same as \cref{def:symmetric-V}. Therefore, the $S$-preserving path-recording oracle $\PRO$ with $d = n^{\omega(1)}$ satisfies that for every $\poly(n)$-query oracle adversary $\adv$ with queries in any order,
    \begin{align*}
    \left| \Pr_{\calO \gets \mu^\Haar_{S^\perp}}  \left[ 1 \gets \adv^{\cO, \cO^\dag} \right] - \Pr_{\gsL \gets \ket{\emptyset}, \gsR \gets \ket{\emptyset}} \left[ 1 \gets \adv^{\PRO, \PRO^\dag} \right] \right| \leq \negl(n).
\end{align*}

Let us describe how to efficiently implement (the extended) $\PRO^\mathrm{L}$ with ancillas. A similar implementation applies to $\PRO^\mathrm{R}$. Together, they give an efficient implementation of the $S$-preserving $\PRO$ and  $\PRO^\dagger$. 
To efficiently implement (the extended) $\PRO^\mathrm{L}$ on input 
$$\ket{x} \ket{L} \ket{R},$$
one prepends a zero-initialized register, creates the uniform superposition over $[d-|\Im(L \cup R)]$ on it controlled on $x \in [d]$, and copies $x$ to it controlled on $x \not\in [d]$. This produces
$$\begin{cases}
    \frac{1}{\sqrt{d - |\Im(L\cup R)|}}\sum_{y \in [d - |\Im(L\cup R)|]} \ket{y} \ket{x} \ket{L} \ket{R} & \text{ if } x \in [d]\\
    \ket{x}\ket{x} \ket{L} \ket{R} & \text{ if } x \not \in [d]
\end{cases}.$$
From the description of $\Im(L\cup R)$, one can efficiently compute the bijection (and its inverse) between $[d - |\Im(L\cup R)|]$ and $\set{y \in [d] : y \not \in \Im(L\cup R)}$. This enables us to first compute the bijection to a new register and then uncompute the value on the original register controlled on $x \in [d]$. This produces
$$\begin{cases}
    \frac{1}{\sqrt{d - |\Im(L\cup R)|}}\sum_{y \in [d],\; y \not \in \Im(L\cup R)} \ket{y}\ket{x} \ket{L} \ket{R}  & \text{ if } x \in [d]\\
    \ket{x}\ket{x} \ket{L} \ket{R}  & \text{ if } x \not \in [d]
\end{cases}.$$
Then, we compute the function that maps
$(y,x,L)$ to $L$ if $x \not \in [d]$, and to $L \cup \set{(x,y)}$ if $x \in [d]$. This produces
$$\begin{cases}
    \frac{1}{\sqrt{d - |\Im(L\cup R)|}}\sum_{y \in [d],\; y \not \in \Im(L\cup R)} \ket{y}\ket{x} \ket{L} \ket{L \cup \set{(x,y)}} \ket{R}  & \text{ if } x \in [d]\\
    \ket{x}\ket{x} \ket{L} \ket{L} \ket{R}  & \text{ if } x \not \in [d]
\end{cases}.$$
Finally, we can use the first and fourth registers to uncompute $x,L$ in the second and third registers by applying the function that maps $(y, L')$ to $(y, L')$ if $y \not \in [d]$, and to $L' \setminus \set{(\wt{x},y)}$ if $y \in [d]$ where $(\wt{x},y)$ is the unique entry in $L'$ with second component $y$. Hence, we arrive at the following state, which is exactly the definition of $\PRO^{\mathrm{L}} \cdot \ket{x}\ket{L}\ket{R}$.
$$\begin{cases}
    \frac{1}{\sqrt{d - |\Im(L\cup R)|}}\sum_{y \in [d],\; y \not \in \Im(L\cup R)} \ket{y} \ket{L \cup \set{(x,y)}} \ket{R} & \text{ if } x \in [d]\\
    \ket{x} \ket{L} \ket{R}  & \text{ if } x \not \in [d]
\end{cases}.$$
We summarize these results as the following lemma.

\begin{lemma}[$S$-preserving path-recording oracle]
\label{lemma:our-PRO}
        Let $d = n^{\omega(1)}$, $S = \Span\set{\ket{x}}_{x \in \zo^n \setminus [d]}$. Let $\PRO$ be the operator defined by extending \cref{def:symmetric-V} by (\ref{eq:extend-PRO-1}), (\ref{eq:extend-PRO-2}), which we call the $S$-preserving path-recording oracle. Then $\PRO,\PRO^\dagger$ can be implemented in $\poly(n)$-time, and for every $\poly(n)$-query oracle adversary $\adv$ with queries in any order,
        \begin{align*}
        \left| \Pr_{\calO \gets \mu^\Haar_{S^\perp}}  \left[ 1 \gets \adv^{\cO, \cO^\dag} \right] - \Pr_{\gsL \gets \ket{\emptyset}, \gsR \gets \ket{\emptyset}} \left[ 1 \gets \adv^{\PRO, \PRO^\dag} \right] \right| \leq \negl(n).
        \end{align*}
\end{lemma}

\section{Quantum Obfuscation}
\label{sec:obfuscation}

In this section, we present the first obfuscation scheme for arbitrary quantum circuits that implement any quantum map. To characterize its security, we introduce in \cref{sec:ideal-obfuscation} the notion of ideal obfuscation for general quantum maps, thus extending the concept of ideal obfuscation that was previously defined only for deterministic classical functions and unitary transformations to a fully general quantum setting. In \cref{sec:invariant-measure}, we propose an ideal functionality that applies to arbitrary quantum maps. In \cref{sec:main-theorem}, we construct our obfuscation scheme and prove that it satisfies the proposed ideal obfuscation security. Furthermore, we establish that this definition of ideal obfuscation implies indistinguishability obfuscation, thereby demonstrating the strength and generality of our construction.

It is well known that every quantum circuit can be efficiently transformed into the following form without loss of generality by deferring measurements and using additional swap gates $\SWAP = \sum_{a,b \in \zo} \ket{a,b}\bra{b,a}$.
\begin{dfn}\label{def:quantum-circuit}
    A quantum circuit is said to have parameter $\parameter = (n,n',m,s)$ if it initializes $(m - n)$ ancillary $\ket{1}$ qubits, takes an $n$-qubit input, applies a sequence of $s$ unitary gates on those qubits, and outputs the last $n'$ qubits. We say that $m$ is the width and $s$ is the size of the circuit. The following objects are associated with a quantum circuit $Q$.
    \begin{itemize}
        \item $U_Q$ is the unitary operator obtained by composing the sequence of unitary gates of $Q$.
        \item $\Gamma_Q = \Tr_{[:-n']} \left[U_Q( 1^{m-n} \ot \cdot )U_Q^\dag\right]$ is the quantum map implemented by $Q$.
    \end{itemize}
\end{dfn}

We can assume without loss of generality that $m \ge n + 2n'$.

\begin{dfn}We define the following classes of quantum circuits.
\begin{align*}
        &\Qarb := \set{Q = (Q_\secparam)_{\secparam \in \bbN} \given 
        Q_\secparam \text{ is a quantum circuit of } \poly(\secparam) \text{ width and } \poly(\secparam) \text{ size}}\\
        &\Qcla := \set{Q \in \Qarb \given 
        \exists \text{ classical function } F_\secparam \text{ such that} \norm{{Q_\secparam}(\cdot) - F_\secparam(\cdot)}_\diamond = \negl(\secparam)}\\
        &\Quni := \set{Q \in \Qarb \given 
        \exists \text{ unitary transformation } \cU_\secparam \text{ such that} \norm{{Q_\secparam}(\cdot) - \cU_\secparam(\cdot)}_\diamond = \negl(\secparam)}
    \end{align*}
\end{dfn}

Our goal would be to obfuscate any quantum circuit in $\Qarb$.
We will work in the classical oracle model, where the obfuscator can output an efficient classical oracle that is accessible to all parties. We start by defining the syntax and basic requirements of a quantum obfuscation scheme.

\begin{dfn}[Quantum Obfuscation]
\label{dfn:obf-syntax-completeness}
 A quantum obfuscation scheme for a class $\cQ$ of quantum circuits in the classical oracle model is a pair of QPT algorithms $(\QObf,\QEval)$ with the following syntax:
\begin{itemize}
    \item $\QObf(1^\secparam, Q) \to (\wt{\psi},\sF)$: The obfuscator takes as input the security parameter $1^\secparam$ and a quantum circuit $Q \in \cQ$, and outputs an obfuscated program specified by an auxiliary quantum state $\wt{\psi}$ and a classical function $\sF$. 
    \item $\QEval^{\sF}(\rho_{\intext},\wt{\psi}) \to \rho_{\out}$: The evaluation algorithm takes as input a quantum input $\rho_{\intext}$ and an auxiliary state $\wt{\psi}$, makes quantum queries to a classical oracle $\sF$ during the evaluation process, and produces a quantum output $\rho_{\out}$.
\end{itemize}
The scheme is required to be \textbf{functionality-preserving}, in the sense that for every $Q \in \cQ$,
$$\E_{(\widetilde{\psi}, \sF) \leftarrow \QObf(1^\lambda, Q)} \norm{\QEval^\sF\left(\cdot, \wt{\psi}\right) - Q(\cdot)}_\diamond = \negl(\secparam).$$
Moreover, the scheme is required to be functionality-preserving with \textbf{reusability}, where the evaluation algorithm $\QEval(\rho_\intext, \wt{\psi}) \to (\rho_\out, \wt{\psi}')$ additionally produces an auxiliary state along with the evaluation output, such that for every $Q \in \cQ$,
$$\E_{(\widetilde{\psi}, \sF) \leftarrow \QObf(1^\lambda, Q)} \norm{\QEval^\sF\left(\cdot, \wt{\psi}\right) - \left(Q(\cdot), \wt{\psi}\right)}_\diamond = \negl(\secparam).$$
\end{dfn}
Note that reusability has been required and achieved either implicitly or explicitly throughout the quantum obfuscation literature. We provide a simple formalization of reusability here, which requires that the evaluation algorithm recover the original auxiliary state $\wt{\psi}$ after evaluation. In terms of security, a standard notion is indistinguishability obfuscation. Below, we state a robust version of it that considers not just functionally equivalent circuits, but also functionally approximately equivalent circuits.

\begin{dfn}[Indistinguishability Obfuscation]\label{def:iO}
    A quantum obfuscation scheme for a class $\cQ$ of quantum circuits in the classical model is said to achieve indistinguishability obfuscation if for every QPT adversary $\adv$ and every $Q, Q' \in \cQ$ with the same parameter (\cref{def:quantum-circuit}) that satisfy $\norm{Q(\cdot) - Q'(\cdot)}_\diamond \le \negl(\secparam)$, it holds that
    \[ \abs{\Pr_{(\widetilde{\psi}, \sF) \leftarrow \QObf(1^\secparam, Q)} \left[1 \gets \adv^\sF\left(\wt{\psi} \right) \right] - \Pr_{(\widetilde{\psi}, \sF) \leftarrow \QObf(1^\secparam, Q')} \left[1 \gets \adv^{\sF}\left(\wt{\psi} \right) \right]} = \negl(\secparam).
    \]
\end{dfn}


\subsection{Ideal Obfuscation}
\label{sec:ideal-obfuscation}

In the literature, ideal obfuscation has only been defined when the computed mapping $\Phi$  is a classical deterministic function \cite{barak2001possibility} or a unitary transformation \cite{huang2025obfuscation}. In order to generalize the strong, simulation-based security notion of ideal obfuscation to arbitrary mappings, we have to reformulate ideal obfuscation in a way that highlights its core characterizations. 

In our new formulation, ideal obfuscation should require that anything that an adversary learns from the obfuscated circuit is computationally indistinguishable from what the adversary learns from interacting with an \emph{ideal functionality} $\Ideal_{\secparam,\parameter,\Phi}$ which depends \emph{only on} the security parameter $\secparam$, the circuit parameter $\parameter$, and the quantum map $\Phi$ implemented by the circuit. Moreover, the ideal functionality should faithfully represent what can be computed generically from a circuit that implements $\Phi$. For example, if the ideal functionality were equipped with a PSPACE oracle, it could compute strictly more than what the original quantum circuit can reveal, leading to a weakened notion of security. Hence, we require that $\Ideal_{\secparam,\parameter,\Phi}$ should be \emph{efficiently implementable} within a negligible statistical error given $\secparam$ and \emph{any quantum circuit} with parameter $\parameter$ that implements $\Phi$. These requirements ensure that the ideal functionality considered captures just as much computational power as what having a circuit would generically provide.

Given the above discussion, we propose the following general definition of ideal obfuscation on top of \cref{dfn:obf-syntax-completeness}. Moreover, we adopt a robust definition, where the quantum circuit $Q$ is allowed to implement the quantum map $\Phi$ of interest up to a negligible statistical error in diamond distance.

\begin{dfn}[Ideal Obfuscation]\label{def:ideal-obfuscation}
    A quantum obfuscation scheme for a class $\cQ$ of quantum circuits in the classical model (\cref{dfn:obf-syntax-completeness}) is said to achieve ideal obfuscation if there exists a family of ideal functionalities $\set{\Ideal_{\secparam,\parameter,\Phi}}_{\secparam,\parameter,\Phi}$ (modeled as an interactive quantum machine) with the following properties.
    \begin{itemize}
        \item Simulation Security: There exists a stateful QPT simulator $\Sim$ such that for every QPT adversary $\adv$ and every quantum circuit $Q \in \cQ$  with input length $n$, output length $n'$, circuit width $m$, and circuit size $s$, \ie with circuit parameter $\parameter = (n,n',m,s)$ implementing a quantum map $\Phi$ (possibly up to a negligible statistical error in diamond distance, \ie $\norm{Q_\secparam(\cdot) - \Phi_\secparam(\cdot)}_{\diamond} \le \negl(\secparam)$), it holds that
        \[ \abs{\Pr_{(\widetilde{\psi}, \sF) \leftarrow \QObf(1^\secparam, Q)} \left[1 \gets \adv^\sF\left(\wt{\psi} \right) \right] - 
        \Pr_{
        (\widetilde{\psi},\sF_\Sim) \leftarrow \Sim^{\Ideal_{\secparam,\parameter,\Phi}}} \left[1 \gets \adv^{\sF_\Sim^{\Ideal_{\secparam,\parameter,\Phi}}}\left(\wt{\psi}\right) \right]} = \negl(\secparam),
        \]
        where $\Sim$ and $\sF_\Sim$ get black-box access to $\Ideal_{\secparam,\parameter,\Phi}$, which only depends on $\secparam,\parameter,\Phi$.
        \item Faithfulness: There exists a QPT algorithm that, when given as input $1^\secparam$ and any circuit $Q \in \cQ$ with parameter $\parameter$ that computes $\Phi$ (possibly up to a negligible statistical error in diamond distance), implements the ideal functionality $\Ideal_{\secparam,\parameter,\Phi}$ with at most $\negl(\secparam)$ statistical error in diamond distance.
    \end{itemize} 
\end{dfn}

For example, the notion of ideal obfuscation of pseudo-deterministic circuits given in \cite{bartusek2023obfuscation,bartusek2024quantum} can be recast as \cref{def:ideal-obfuscation} with the following ideal functionality.

\begin{idealmodel}{Ideal Functionality $\Ideal^{\mathsf{pseudo\text{-}det}}_{\secparam,\parameter,\Phi}$}
    \begin{itemize}
        \item Set $F(\cdot)$ as the classical function closest to $\Phi(\cdot)$ in diamond distance. 
        \item On initialization, output $(1^\secparam, 1^\parameter)$.
        \item On query $\ket{x,y} \in \bbC^{2^n} \ot \bbC^{2^{n'}}$, apply $F$ and output the result $\ket{x,y \oplus F(x)}$.
    \end{itemize}
\end{idealmodel}
For any pseudo-deterministic circuit computing $\Phi$, the corresponding $F$ is well-defined. A simulator with access to this ideal functionality would obtain $(1^\secparam, 1^\parameter)$ at the beginning and have quantum query access to $F$, which is exactly the definition in \cite{bartusek2023secure,bartusek2024quantum}. Likewise, the ideal obfuscation of unitary circuits in \cite{huang2025obfuscation} can be phrased in terms of \cref{def:ideal-obfuscation}.
For convenience, here we state the security of \cite{huang2025obfuscation} in the original formulation.

\begin{thm}[\cite{huang2025obfuscation}, Obfuscation of unitaries\footnote{The scheme of \cite{huang2025obfuscation} is functionality-preserving with reusability (\cref{dfn:obf-syntax-completeness}), which can be derived directly from their proof that the scheme is functionality-preserving (in particular, they showed that $\QObf$ preserves the functionality with negligible statistical error) together with their proof of reusability (which applies whenever the functionality computes a unitary transformation with negligible statistical error).}]
\label{thm:obf-unitary}
    There exists an ideal obfuscation scheme $(\QObf, \QEval)$ for the class $\Quni$ of unitary quantum circuits in the classical oracle model,
    assuming the existence of post-quantum one-way functions, in the following sense: There exists a stateful QPT simulator $\Sim$ such that for every QPT adversary $\adv$ and every quantum circuit $Q \in \Quni$ with input and output length $n$, circuit width $m$, and circuit size $s$, \ie with parameter $\parameter = (n,n,m,s)$ that implements a unitary transformation $\cU$ up to a negligible statistical error in diamond distance, it holds that
    {\small
    \[
    \abs{\Pr_{(\widetilde{\psi}, \sF) \leftarrow \QObf(1^\secparam, Q)} \left[1 \gets \adv^\sF\left(\wt{\psi} \right) \right] - 
    \Pr_{(\widetilde{\psi}, \sF_\Sim) \leftarrow \Sim(1^\secparam, 1^\parameter)} \left[1 \gets \adv^{\sF_{\Sim}^{\cU, \cU^\dag}}\left(\wt{\psi}\right) \right]} = \negl(\secparam),
    \]
    }where $\sF_\Sim$ makes black-box queries to $\cU$ and its inverse $\cU^\dag$ in a fixed order\footnote{In particular, $\sF_\Sim$ makes its odd-numbered queries to $\cU$ and even-numbered queries to $\cU^\dag$.}.
\end{thm}

It is also direct to see that any scheme satisfying our definition of ideal obfuscation (\cref{def:ideal-obfuscation}) automatically achieves indistinguishability obfuscation (\cref{def:iO}). This is because given any two functionally approximately equivalent circuits $Q$ and $Q'$ with the same circuit parameter $\parameter$, both circuits implement the quantum mapping $\Phi:=\Gamma_Q$ up to negligible errors in diamond distance. Ideal obfuscation security implies that both the obfuscation of $Q$ and the obfuscation of $Q'$ are computationally indistinguishable from the same ideal situation, and thus the two obfuscations are computationally indistinguishable.

\subsection{Ideal Functionality}
\label{sec:invariant-measure}
We have to identify a suitable definition of ideal functionality $\Ideal_{\secparam,\parameter,\Phi}$ for an arbitrary quantum map $\Phi$. This involves various considerations. For example, a natural attempt would be to define $\Ideal_{\secparam,\parameter,\Phi}$ as providing $\secparam,\parameter$ and query access to $\Phi$. However, ideal obfuscation can never be achieved under this definition of ideal functionality when $\Phi$ is a unitary transformation, because any obfuscation of $\Phi$ can be used to implement $\Phi^{-1}$ as well \cite{huang2025obfuscation}, which is not implementable in general using the proposed ideal functionality. Therefore, one would like to include the inverse of $\Phi$ in the ideal functionality, but the inverse is not well-defined for a general quantum map $\Phi$. Moreover, the ideal functionality $\Ideal_{\secparam,\parameter,\Phi}$ is required to depend only on $\secparam,\parameter,\Phi$ and be efficiently implementable from circuits for $\Phi$.

To overcome the aforementioned difficulties and define $\Ideal_{\secparam,\parameter,\Phi}$ for an arbitrary quantum map $\Phi$ in a canonical manner, we consider the following notion of (the topological space of) \emph{unitary extensions} of a quantum map $\Phi$ and define a unique measure on it.

\begin{dfn}[Unitary extension]
\label{def:unitary-extension}
    Given a quantum map $\Phi: \Den{n} \to \Den{n'}$ and an integer $m \ge n + 2n'$, we define the set of $m$-qubit unitary extensions of $\Phi$ as
    $$\cU(2^{m})|_\Phi := \set{V \in \cU(2^{m}) \;|\; \Tr_{[:-n']} V(1^{m-n} \ot \cdot)V^\dag = \Phi(\cdot)}$$
    consisting of all $m$-qubit unitary operators that compute $\Phi$ when the ancillary qubits are all initialized to the $\ket{1}$ state and a partial trace is taken at the end. We equip $\cU(2^{m})|_\Phi$ with the induced topology under the inclusion map $\cU(2^{m})|_\Phi \xhookrightarrow{} \cU(2^{m})$.
\end{dfn}

It is clear that a quantum map $\Phi$ can be efficiently computed given black-box access to any of its unitary extensions $V \in \cU(2^{m})|_\Phi$. By our convention (\cref{def:quantum-circuit}), a quantum circuit $Q$ with parameter $\parameter = (n,n',m,s)$ naturally yields an efficient unitary extension $U_Q \in \cU(2^m)|_{\Gamma_Q}$ of the quantum map $\Gamma_Q$ that $Q$ implements.

We would like to define an invariant probability measure over the set of $m$-qubit unitary extensions $\cU(2^{m})|_\Phi$ of a quantum map $\Phi$, where the measure depends solely on $m$ and $\Phi$. Such a probability measure will provide a suitable definition of $\Ideal_{\secparam,\parameter,\Phi}$.

For the following lemma, recall the definition of $\cU(S)$ for a subspace $S \subseteq \bbC^{2^m}$ (\cref{sec:topo-haar}).

\begin{lemma}\label{lemma:closed-under-composition}
    The set $\cU(2^{m})|_\Phi$ is non-empty and closed under
    \begin{itemize}
        \item Left multiplication by $(U \ot I_{n'})$ for any $(m-n')$-qubit unitary $U \in \cU(2^{m-n'})$.
        \item Right multiplication by any $m$-qubit unitary $U' \in \cU(\Span\set{\ket{x}}_{x \in [2^{m} - 2^n]})$.
    \end{itemize}
\end{lemma}
\begin{proof}
The non-emptiness follows directly from the well-known fact that $\Phi$ has a Stinespring representation for any $m \ge n + 2n'$ (\eg see \cite{watrous2025understanding}).
For any $V\in\cU(2^m)|_\Phi$, we know that
$$\Gamma_V(\cdot) = \Tr_{[:-n']}[V(1^{m-n}\ot \cdot)V^\dag] = \Phi(\cdot)$$
Then for $U \in \cU(2^{m-n})$, we have
\begin{align*}
    \Gamma_{(U \ot I_{n'}) V}(\cdot) &= \Tr_{[:-n']}[(U \ot I_{n'}) V(1^{m-n}\ot \cdot)V^\dag (U \ot I_{n'})^\dag]\\
    &= \Tr_{[:-n']}[V(1^{m-n}\ot \cdot)V^\dag] = \Phi(\cdot)
\end{align*}
using the cyclic property of partial trace in the second equality. Thus, $(U \ot I_{n'}) V \in \cU(2^m)|_\Phi$. Also, for $U' \in \cU(\Span \set{\ket{x}}_{x \in [2^m - 2^n]})$, we have
\begin{align*}
    \Gamma_{V U'}(\cdot) &= 
    \Tr_{[:-n']}[V U' (1^{m-n}\ot \cdot) U'^\dag V^\dag] \\
    &= \Tr_{[:-n']}[V(1^{m-n}\ot \cdot)V^\dag] = \Phi(\cdot)
\end{align*}
using the fact that $(1^{m-n} \ot \cdot)$ always lies in the subspace $\Span \set{\ket{x}}_{x \in \zo^m \setminus [2^m - 2^n]}$ on which $U'$ acts as the identity map. Thus, $V U' \in \cU(2^m)|_\Phi$.
\end{proof}

\begin{lemma}
\label{lemma:our-uniform-distribution}
    There exists a unique probability measure $\mu^{\unif}_{\Phi}$ on the Borel subsets of $\cU(2^{m})|_\Phi$ such that for every Borel measurable set $A \subseteq \cU(2^{m})|_\Phi$, $(m-n')$-qubit unitary $U \in \cU(2^{m-n'})$, and $m$-qubit unitary $U' \in \cU(\Span\set{\ket{x}}_{x \in [2^{m} - 2^n]})$, the probability measure satisfies
    $$\mu^{\unif}_{\Phi}((U \ot I_{n'}) A) = \mu^{\unif}_{\Phi}(A U') =  \mu^{\unif}_{\Phi}(A).$$
We sometimes write $\mu^\unif_{\Phi}(2^m)$ instead of $\mu^\unif_\Phi$ to stress that it is a measure over $\cU(2^m)|_\Phi$. We also identify $\mu^\unif_{\Phi}(2^m)$ with its pushforward measure on $\cU(2^{m})$ under the inclusion map.

\end{lemma}
\noindent 
\begin{proof}
For $V \in \cU(2^m)|_\Phi$, consider the continuous map
$$f_V: \cU(2^{m-n'}) \times \cU(\Span\set{\ket{x}}_{x \in [2^m - 2^n]}) \to \cU(2^m)|_\Phi$$
given by $$f_V(U, U') = (U \ot I_{n'}) V U'.$$
The map $f_V$ is well-defined according to \cref{lemma:closed-under-composition}.
Let $\mu_{V}$ be the pushforward measure of $\mu^\Haar_{2^{m-n'}} \times \mu^\Haar_{\Span\set{\ket{x}}_{x \in [2^m - 2^n]}}$ onto $\cU(2^m)|_\Phi$ under $f_V$. In other words, one can sample from $\mu_V$ by first sampling $U \sim \mu^\Haar_{2^{m-n'}}$, $U' \sim \mu^\Haar_{\Span\set{\ket{x}}_{x \in [2^m - 2^n]}}$ and then outputting $(U \ot I_{n'}) V U'$.
For every measurable set $A \subseteq \cU(2^m)|_\Phi$, and $(m-n')$-qubit unitary $\wt{U} \in \cU({2^{m-n'}})$, and $m$-qubit unitary $\wt{U'} \in \cU(\Span\set{\ket{x}}_{x \in [2^m - 2^n]})$, we have
\begin{align*}
    \mu_V\left((\wt{U} \ot I_{n'}) A\right) &= \int 1_{(\wt{U} \ot I_{n'}) A} \left(W\right)\; \mu_V (dW)\\
    &= \int 1_{(\wt{U} \ot I_{n'}) A} \left(f_V(U,U')\right) \left(\mu^\Haar_{2^{m-n'}} \times \mu^\Haar_{\Span\set{\ket{x}}_{x \in [2^m - 2^n]}}\right) \left(dU, dU'\right)\\
    &= \iint 1_{A} \left((\wt{U}^{-1} \ot I_{n'}) f_V(U,U')\right) \mu^\Haar_{2^{m-n'}}(dU)\;  \mu^\Haar_{\Span\set{\ket{x}}_{x \in [2^m - 2^n]}}(dU')\\
    &= \iint 1_{A} \left((\wt{U}^{-1} \ot I_{n'}) f_V(\wt{U} U,U')\right) \mu^\Haar_{2^{m-n'}}(d (\wt{U}U))\;  \mu^\Haar_{\Span\set{\ket{x}}_{x \in [2^m - 2^n]}}(dU')\\
    &= \iint 1_{A} \left(f_V( U,U')\right) \mu^\Haar_{2^{m-n'}}(d U)\;  \mu^\Haar_{\Span\set{\ket{x}}_{x \in [2^m - 2^n]}}(dU') = \mu_V(A)
\end{align*}
where the second equality is due to the product distribution and the fact that $f_V(U,U') \in (\wt{U} \ot I_{n'}) A$ if and only if $(\wt{U}^{-1} \ot I_{n'}) f_V(U,U') \in A$. The third equality uses a change of variables from $U$ to $\wt{U} U$. The fourth equality is by definition of $f_V$ and the left invariance property of $\mu^\Haar_{2^{m-n'}}$.
\begin{align*}
    \mu_V(A \wt{U'}) &= \int 1_{A \wt{U'}} \left(W\right)\; \mu_V (dW)\\
    &= \int 1_{A\wt{U'}} \left(f_V(U,U')\right) \left(\mu^\Haar_{2^{m-n'}} \times \mu^\Haar_{\Span\set{\ket{x}}_{x \in [2^m - 2^n]}}\right) \left(dU, dU'\right)\\
    &= \iint 1_{A} \left(f_V(U,U') \wt{U'}{}^{-1}\right) \mu^\Haar_{2^{m-n'}}(dU)\;  \mu^\Haar_{\Span\set{\ket{x}}_{x \in [2^m - 2^n]}}(dU')\\
    &= \iint 1_{A} \left( f_V(U,U' \wt{U'}) \wt{U'}{}^{-1}\right) \mu^\Haar_{2^{m-n'}}(d U)\;  \mu^\Haar_{\Span\set{\ket{x}}_{x \in [2^m - 2^n]}}(d(U' \wt{U'}))\\
    &= \iint 1_{A} \left(f_V( U,U')\right) \mu^\Haar_{2^{m-n'}}(d U)\;  \mu^\Haar_{\Span\set{\ket{x}}_{x \in [2^m - 2^n]}}(dU') = \mu_V(A)
\end{align*}
where we use the equivalence between $f_V(U,U') \in A \wt{U'}$ and $f_V(U,U')\wt{U'}{}^{-1} \in A$ and the right invariance property of $\mu^\Haar_{\Span\set{\ket{x}}_{x \in [2^m - 2^n]}}$. Therefore, $\mu_V$ is a probability measure that satisfies the required invariance properties.

Next, for any $V,V' \in \cU(2^m)|_\Phi$, we will show that $\mu_V = \mu_{V'}$. By \cref{thm:continuity-of-dilation}, there exists $\wt{U} \in \cU(2^{m-n'})$ such that 
\begin{equation}
\label{eq:invariance-1}
    V' (1^{m-n} \ot \cdot) V'^\dag = (\wt{U} \ot I_{n'}) V (1^{m-n} \ot \cdot) V^\dag (\wt{U}^\dag \ot I_{n'}).
\end{equation}
Set $\wt{U'} = V^\dag (\wt{U}^\dag \ot I_{n'}) V' \in \cU(2^m)$. Then for every $\ket{x} \in \bbC^{2^m}$ where $x \not\in [2^m - 2^n]$, we can write $\ket{x} = \ket{1^{m-n},y}$ where $y \in \zo^n$, and from (\ref{eq:invariance-1}) we can derive that
$$\wt{U'} \ketbra{x} \wt{U'}^\dag = \left(V^\dag (\wt{U}^\dag \ot I_{n'}) V'\right) \left(1^{m-n} \ot \ketbra{y} \right) \left(V'^\dag (\wt{U} \ot I_{n'}) V\right) = 1^{m-n} \ot \ketbra{y} = \ketbra{x}$$
Therefore, $\wt{U'} \in \cU(\Span\set{\ket{x}}_{x \in [2^m - 2^n]})$. By the definition of $\wt{U'}$, we have $V' = (\wt{U} \ot I_{n'}) V \wt{U'}$. 
For any measurable set $A \subseteq \cU(2^m)|_\Phi$, we have
\begin{align*}
    \mu_{V'}(A) =& \iint 1_{A} \left(f_{V'}( U,U')\right) \mu^\Haar_{2^{m-n'}}(d U)\;  \mu^\Haar_{\Span\set{\ket{x}}_{x \in [2^m - 2^n]}}(dU')\\
    =& \iint 1_{A} \left((U \ot I_{n'}) (\wt{U} \ot I_{n'}) V \wt{U'} U'\right) \mu^\Haar_{2^{m-n'}}(d U)\;  \mu^\Haar_{\Span\set{\ket{x}}_{x \in [2^m - 2^n]}}(dU')\\
    =& \iint 1_{A} \left(f_{V}( U \wt{U},\wt{U'} U')\right) \mu^\Haar_{2^{m-n'}}(d U)\;  \mu^\Haar_{\Span\set{\ket{x}}_{x \in [2^m - 2^n]}}(dU')\\
    =& \iint 1_{A} \left(f_{V}( U , U')\right) \mu^\Haar_{2^{m-n'}}(d (U\wt{U}^{-1}))\;  \mu^\Haar_{\Span\set{\ket{x}}_{x \in [2^m - 2^n]}}(d(\wt{U'}{}^{-1}U'))\\
    =& \iint 1_{A} \left(f_{V}( U,U')\right) \mu^\Haar_{2^{m-n'}}(d U)\;  \mu^\Haar_{\Span\set{\ket{x}}_{x \in [2^m - 2^n]}}(dU') = \mu_V(A)
\end{align*}
where the fourth equality uses a change of variables and the next equality uses the invariance properties of $\mu^\Haar_{2^{m-n'}}$ and $ \mu^\Haar_{\Span\set{\ket{x}}_{x \in [2^m - 2^n]}}$. Thus $\mu_V = \mu_{V'}$. We now define $\mu^\unif_{\Phi}$ to be $\mu_V$ for any choice of $V \in \cU(2^m)_{\Phi}$, which is well-defined. This proves the existence of $\mu^\unif_{\Phi}$.

Let $\nu$ be any probability measure on $\cU(2^m)_{\Phi}$ satisfying these properties. Then for any measurable set $A \subseteq \cU(2^m)_{\Phi}$, we have
\begin{align*}
    \mu^\unif_{\Phi}(A) &= \int 1_A(W) \mu^\unif_{\Phi}(dW)\\ 
    &= \iint 1_A(W) \mu_V(dW)\nu(dV)\\
    &= \iint 1_A(f_V(U,U')) \left(\mu^\Haar_{2^{m-n'}} \times \mu^\Haar_{\Span{\set{\ket{x}}_{x \in [2^m - 2^n]}}}\right) (dU, dU')\;\nu(dV)\\
    &= \iint 1_A((U \ot I_{n'})V U') \left(\mu^\Haar_{2^{m-n'}} \times \mu^\Haar_{\Span{\set{\ket{x}}_{x \in [2^m - 2^n]}}}\right) (dU, dU')\;\nu(dV)\\
    &= \iint 1_A((U \ot I_{n'})V U') \nu(dV)\left(\mu^\Haar_{2^{m-n'}} \times \mu^\Haar_{\Span{\set{\ket{x}}_{x \in [2^m - 2^n]}}}\right) (dU, dU')\\
    &= \iint 1_A(V) \nu (dV) \left(\mu^\Haar_{2^{m-n'}} \times \mu^\Haar_{\Span{\set{\ket{x}}_{x \in [2^m - 2^n]}}}\right) (dU, dU')\\
    &= \int \nu(A) \left(\mu^\Haar_{2^{m-n'}} \times \mu^\Haar_{\Span{\set{\ket{x}}_{x \in [2^m - 2^n]}}}\right) (dU, dU') = \nu(A)
    \end{align*}
    The second equality follows because $\nu$ is a probability measure and $\mu^\unif_{\Phi} = \mu_V$ for any $V \in \cU(2^m)_{\Phi}$. The third equality follows because $\mu_V$ is the pushforward measure of $\mu^\Haar \times \mu^\Haar$ under $f_V$. The fifth equality follows from Fubini's theorem. The sixth equality follows from the invariance properties of $\nu$.
    We conclude that $\nu = \mu^\unif_{\Phi}$. This proves the uniqueness of such a probability measure.
\end{proof}

\begin{rmk}
\label{rmk:sampling-uniform-distribution}
    The proof of \cref{lemma:our-uniform-distribution} shows that $\mu^\unif_{\Phi}(2^m)$ can be sampled in various ways. In particular, it can be sampled by first sampling $U \sim \mu^\Haar_{2^{m-n'}}$, $U' \sim \mu^\Haar_{\Span\set{\ket{x}}_{x \in [2^m - 2^n]}}$ and then outputting the unitary  $(U \ot I_{n'}) V U'$ using any $V \in \cU(2^m)|_\Phi$.
\end{rmk}
For arbitrary quantum mappings, we propose the following ideal functionality $\Ideal_{\secparam,\parameter,\Phi}$, whose description depends only on $\secparam,\parameter,\Phi$ according to \cref{lemma:our-uniform-distribution}.

\begin{figure}[H]
\begin{idealmodel}{Ideal Functionality $\Ideal_{\secparam,\parameter,\Phi}$.}
    \begin{itemize}[leftmargin=*]
        \item On initialization, let $\parameter = (n,n',m,s)$, sample $W \gets \mu^\unif_\Phi(2^{\secparam+m})$, and output $(1^\secparam, 1^\parameter)$.
        \item On query $(b \in \set{\pm1}, \rho \in \Den{\secparam+m})$, apply $W^b$ to $\rho$ and output the resulting state.
    \end{itemize}
\end{idealmodel}
\caption{Ideal functionality for a general quantum mapping $\Phi$.}
\label{fig:our-ideal-functionality}
\end{figure}

\subsection{Construction and Main Theorem}
\label{sec:main-theorem}

This section is dedicated to proving the following main theorem.

\begin{theorem}[Obfuscation of arbitrary quantum circuits]\label{thm:obfuscate-all}
    There exists an ideal obfuscation scheme for the class $\Qarb$ of arbitrary quantum circuits in the classical oracle model satisfying \cref{def:ideal-obfuscation} under the ideal functionality $\set{\Ideal_{\secparam,\parameter,\Phi}}_{\secparam,\parameter,\Phi}$ proposed in \cref{sec:invariant-measure}, assuming the existence of post-quantum one-way functions.
\end{theorem}

\begin{Construction}{Ideal obfuscation of arbitrary quantum circuits.}
\label{construction:obfuscation-of-arbitrary}
\begin{itemize}[leftmargin=*]
    \item Ingredients:
    \begin{itemize}
        \item An ideal obfuscation scheme $(\QObf_{\unitary}, \QEval_{\unitary})$ for $\Quni$. (\cref{thm:obf-unitary})
        \item A strong pseudorandom unitary $\standardPRU$. (\cref{thm:PRU-exists})
        \item A subspace-preserving strong pseudorandom unitary $\spPRU$. (\cref{thm:spsPRU})
    \end{itemize}
    \item $\QObf(1^\secparam, Q)$ for a quantum circuit $Q$ with parameter $\parameter = (n,n',m,s)$:
    \begin{enumerate}
        \item Set $m' = \secparam + m$, dimension $d = 2^{m'} - 2^n$ and subspace $S = \Span\set{\ket{x}}_{x \in \zo^{m'} \setminus [d]}$.
        \item Sample a random key $k \gets \cK$ for the $m'$-qubit $S$-preserving $\spPRU$.
        \item Sample a random key $k' \gets \cK'$ for the $(m'-n')$-qubit $\standardPRU$.
        \item Create a quantum circuit for $\CTRL_{1^\secparam}\text{-}U_Q$, which applies $U_Q$ on qubits $(\secparam + [m])$ controlled on the first $\secparam$ qubits being all $\ket{1}$. This can be efficiently done by replacing each unitary gate $G$ in $Q$ with $\CTRL_{1^\secparam}\text{-}G$, which has an efficient circuit implementation.
        \item Create the unitary quantum circuit $Q'_{k,k'} = (\standardPRU_{k'} \ot I_{n'}) \circ (\CTRL_{1^\secparam}\text{-}U_Q) \circ \spPRU_k$ that has $m'$ input qubits and $m'$ output qubits.
        \item Compute $(\wt{\psi},\sF) \gets \QObf_{\unitary}(1^\secparam, Q'_{k,k'})$ and output $(\wt{\psi}, \sF)$.
    \end{enumerate}
        \item $\QEval^{\sF}(\rho, \wt{\psi})$ for a quantum input $\rho$ with $n$ qubits:
        \begin{enumerate}
            \item Obtain $\rho' \gets \QEval_{\unitary}^\sF\big(1^{m'-n} \ot \rho,\; \wt{\psi}\big)$ along with a recovered auxiliary state $\wt{\psi}'$.
            \item Output the state on the last $n'$ qubits of $\rho'$, which is $\Tr_{[:-n']}(\rho')$.
        \end{enumerate}
    \end{itemize}
\end{Construction}

\begin{proof}
    We show that Construction \ref{construction:obfuscation-of-arbitrary} is an ideal quantum obfuscation scheme for $\Qarb$.
    By construction, $\QObf$ and $\QEval$ are efficient algorithms. To prove that the scheme is functionality-preserving, 
    we start with the property that $\spPRU$ is $S$-preserving for $S = \Span\set{\ket{x}}_{x \in \zo^{m'} \setminus [2^{m'} - 2^n]} = \Span \set{\ket{1^{m'-n}, y}}_{y \in \zo^n}$, so for every $k \in \cK$ we have
    $$\spPRU_k (1^{m'-n} \ot \cdot) \spPRU_k^\dag = (1^{m'-n} \ot \cdot)$$
    and therefore by the cyclicity of partial trace and the definition of $Q$, we have for every $k,k'$,
    \begin{align}
        & \Tr_{[:-n']} \left[Q'_{k,k'}(1^{m'-n} \ot \cdot)\right] \nonumber \\
        =& \Tr_{[:-n']} \left[(\standardPRU_{k'} \ot I_{n'}) (\CTRL_{1^\secparam}\text{-}U_Q) (1^{m'-n} \ot \cdot) (\CTRL_{1^\secparam}\text{-}U_Q^\dag) (\standardPRU_{k'}^\dag \ot I_{n'}) \right] \nonumber \\
        =& \Tr_{[:-n']} \left[(\standardPRU_{k'} \ot I_{n'}) \left(1^\secparam \ot U_Q(1^{m-n} \ot \cdot)U_Q^\dag\right) (\standardPRU_{k'}^\dag \ot I_{n'}) \right] \nonumber \\
        =& \Tr_{[:-n']} \left[\left(1^\secparam \ot U_Q(1^{m-n} \ot \cdot)U_Q^\dag\right)\right] = Q(\cdot)
        \label{eq:completeness-proof-1}
    \end{align}
    Next, by the functionality-preserving property of $(\QObf_{\unitary}, \QEval_{\unitary})$ and the fact that $Q'_{k,k'}$ is defined to implement a unitary transformation, we see that for every $k,k'$,
    \begin{equation*}
        \E_{(\wt{\psi},\sF) \gets \QObf_{\unitary}(1^\secparam, Q'_{k,k'})} \norm{\QEval_{\unitary}^\sF \left(1^{m'-n} \ot \cdot , \wt{\psi}\right) - Q'_{k,k'}(1^{m'-n} \ot \cdot)}_\diamond = \negl(\secparam)
        \label{eq:apply-functionality-preserving-of-obf-unitary}
    \end{equation*}
    Using the fact that the norm is contractive under partial trace and plugging in (\ref{eq:completeness-proof-1}), we get
    \begin{equation*}
        \E_{(\wt{\psi},\sF) \gets \QObf_{\unitary}(1^\secparam, Q'_{k,k'})} \norm{\Tr_{[:-n']}\left[\QEval_{\unitary}^\sF \left(1^{m'-n} \ot \cdot , \wt{\psi}\right)\right] - Q(\cdot)}_\diamond = \negl(\secparam)
    \end{equation*}
    for every $k,k'$. By averaging over $k,k'$ and using the definition of $\QObf, \QEval$, we obtain the functionality-preserving property 
    \begin{equation*}
        \E_{(\wt{\psi},\sF) \gets \QObf(1^\secparam,Q)} \norm{\QEval\left(\cdot , \wt{\psi}\right) - Q(\cdot)}_\diamond = \negl(\secparam).
    \end{equation*}
    
    Note that $(\QObf_\unitary,\QEval_\unitary)$ satisfies reusability, which roughly says that the original auxiliary state is successfully recovered after running $\QEval_\unitary$. By our construction, $\QEval^{\sF}(\cdot,\wt{\psi})$ internally executes $\QEval_\unitary^{\sF}(\cdot,\wt{\psi})$, where $(\wt{\psi},\sF)$ are generated by $\QObf_\unitary$. Therefore, the original auxiliary state is also successfully recovered after running $\QEval^\sF(\cdot, \wt{\psi})$. 
    One can directly augment the arguments above for proving functionality-preserving to derive that the scheme $(\QObf,\QEval)$ is also functionality-preserving with reusability.

    Next, we prove the security of the scheme. Define $\Phi^*(\cdot):=Q(\cdot)$. Denote the parameter of $Q_{k,k'}'$ as $\parameter' = (m',m',w,\ell)$ where its width $w$ and size $\ell$ are polynomials in $\secparam,n,n',m,s$ and do not depend on $k,k'$. The security of $\QObf_{\unitary}$ (\cref{thm:obf-unitary}) implies that there is a simulator $\Sim$ such that for every QPT adversary $\adv$ and every $k,k'$,
    \begin{align*}
        \abs{\Pr_{(\widetilde{\psi}, \sF) \leftarrow \QObf_{\unitary}(1^\secparam, Q'_{k,k'})} \left[1 \gets \adv^\sF\left(\wt{\psi} \right) \right] - 
    \Pr_{(\widetilde{\psi}, \sF_\Sim) \leftarrow \Sim(1^\secparam, 1^{\parameter'})} \left[1 \gets \adv^{\sF_{\Sim}^{Q'_{k,k'}, Q_{k,k'}'^{-1}}}\left(\wt{\psi}\right) \right]} = \negl(\secparam)
    \end{align*}
    because $Q'_{k,k'}$ implements a unitary transformation by construction. By averaging over $k,k'$,
    \begin{align*}
        \abs{\Pr_{(\widetilde{\psi}, \sF) \leftarrow \QObf(1^\secparam, Q)} \left[1 \gets \adv^\sF\left(\wt{\psi} \right) \right] - 
    \Pr_{\substack{k\gets \cK,\; k' \gets \cK'\\
    (\widetilde{\psi}, \sF_\Sim) \leftarrow \Sim(1^\secparam, 1^{\parameter'})}} \left[1 \gets \adv^{\sF_{\Sim}^{Q'_{k,k'}, Q_{k,k'}'^{-1}}}\left(\wt{\psi}\right) \right]} = \negl(\secparam).
    \end{align*}
    Next, by the pseudorandomness of $\standardPRU$ and $\spPRU$, forward and inverse queries to $$Q'_{k,k'} = (\standardPRU_{k'} \ot I_{n'}) \circ (\CTRL_{1^\secparam}\text{-}U_Q) \circ \spPRU_k \text{\;,\; for random } k\gets \cK,\; k'\gets \cK'$$
    are computationally indistinguishable from forward and inverse queries to
    \begin{equation}
        \widehat{Q} = (R' \ot I_{n'}) \circ (\CTRL_{1^\secparam}\text{-}U_Q) \circ R \text{\;,\; for random } R' \sim \mu^\Haar_{2^{m'-n'}},\; R \sim \mu^\Haar_{S^\perp}
        \label{eq:randomized-unitary}
    \end{equation}
    where $S = \Span\set{\ket{x}}_{x \in \zo^{m'} \setminus [2^{m'} - 2^n]}$. In other words,
    \begin{align*}
        \abs{\Pr_{(\widetilde{\psi}, \sF) \leftarrow \QObf(1^\secparam, Q)} \left[1 \gets \adv^\sF\left(\wt{\psi} \right) \right] - 
    \Pr_{\substack{R \sim \mu^\Haar_{S^\perp},\;R' \sim \mu^\Haar_{2^{m'-n'}}\\
    (\widetilde{\psi}, \sF_\Sim) \leftarrow \Sim(1^\secparam, 1^{\parameter'})}} \left[1 \gets \adv^{\sF_{\Sim}^{\widehat{Q},\widehat{Q}^{-1}}}\left(\wt{\psi}\right) \right]} = \negl(\secparam).
    \end{align*}
    To analyze the distribution of $\widehat{Q}$, notice that
    \begin{align*}
        \Gamma_{\CTRL_{1^\secparam}\text{-}U_Q}(\cdot) &= \Tr_{[:-n']} \left[\left(\CTRL_{1^\secparam}\text{-}U_Q\right) (1^{m'-n} \ot \cdot) \left(\CTRL_{1^\secparam}\text{-}U_Q^\dag\right)\right]\\
        &= \Tr_{[:-n']}\left[ \left(1^{\secparam},\; U_Q (1^{m-n} \ot \cdot) U_Q^\dag\right)\right]\\
        &= \Tr_{[:-n']} \left[ U_Q (1^{m-n} \ot \cdot) U_Q^\dag\right] = \Phi^*(\cdot).
    \end{align*}
    Thus,  $\left(\CTRL_{1^\secparam}\text{-}U_Q\right) \in \cU(2^{m'})|_{\Phi^*}$, and  we have $\widehat{Q} \sim \mu^{\unif}_{\Phi^*}$ by \cref{rmk:sampling-uniform-distribution}. Therefore, we get
    \begin{align*}
        \abs{\Pr_{(\widetilde{\psi}, \sF) \leftarrow \QObf(1^\secparam, Q)} \left[1 \gets \adv^\sF\left(\wt{\psi} \right) \right] - 
    \Pr_{\substack{W \gets \mu^\unif_{\Phi^*} \\
    (\widetilde{\psi}, \sF_\Sim) \leftarrow \Sim(1^\secparam, 1^{\parameter'})}} \left[1 \gets \adv^{\sF_{\Sim}^{W,W^\dag}}\left(\wt{\psi}\right) \right]} = \negl(\secparam).
    \end{align*}
    Now recall that  $\Ideal_{\secparam,\parameter,\Phi^*}$ samples $W \gets \mu^\unif_{\Phi^*}(2^{\secparam+m})$, outputs $(1^\secparam,1^\parameter)$, and provides oracle access to $W,W^\dag$. Hence, we can construct a QPT simulator $\Sim'$ for $(\QObf,\QEval)$ that, given an input $(1^\secparam, 1^\parameter)$ from the ideal functionality, runs $\Sim(1^\secparam, 1^{\parameter'})$. Then we obtain
    \begin{align*}
        \abs{\Pr_{(\widetilde{\psi}, \sF) \leftarrow \QObf(1^\secparam, Q)} \left[1 \gets \adv^\sF\left(\wt{\psi} \right) \right] - 
    \Pr_{(\widetilde{\psi}, \sF_\Sim) \leftarrow \Sim'{}^{\Ideal_{\secparam,\parameter,\Phi^*}}} \left[1 \gets \adv^{\sF_{\Sim}^{\Ideal_{\secparam,\parameter,\Phi^*}}}\left(\wt{\psi}\right) \right]} = \negl(\secparam).
    \end{align*}
    This proves simulation security when $Q$ perfectly implements the quantum mapping $\Phi$ of interest, \ie when $\Phi = \Phi^*$. To prove the robust version of simulation security, consider the case where $Q$ implements $\Phi$ up to a negligible error $\eps$, \ie $\norm{Q(\cdot) - \Phi(\cdot)}_\diamond \le \eps = \negl(\secparam)$. In this case, it remains to show that
    \begin{align*}
        \abs{\Pr_{(\widetilde{\psi}, \sF_\Sim) \leftarrow \Sim'{}^{\Ideal_{\secparam,\parameter,\Phi^*}}} \left[1 \gets \adv^{\sF_{\Sim}^{\Ideal_{\secparam,\parameter,\Phi^*}}}\left(\wt{\psi}\right) \right] - 
    \Pr_{(\widetilde{\psi}, \sF_\Sim) \leftarrow \Sim'{}^{\Ideal_{\secparam,\parameter,\Phi}}} \left[1 \gets \adv^{\sF_{\Sim}^{\Ideal_{\secparam,\parameter,\Phi}}}\left(\wt{\psi}\right) \right]} = \negl(\secparam).
    \end{align*}
    Since $\Sim', \sF_\Sim, \adv$ run in polynomial time, it suffices to show that for any polynomial-query distinguisher $\cD$, the ideal functionalities $\Ideal_{\secparam,\parameter,\Phi^*}$ and $\Ideal_{\secparam,\parameter,\Phi}$ can only be distinguished with negligible advantage. In other words, it suffices to show that for any polynomial-query distinguisher $\cD$,
    \begin{align*}
        \abs{\Pr_{W_{0} \gets \mu^\unif_{\Phi^*}} \left[ 1 \gets \cD^{W_0, W_0^\dag}(1^\secparam,1^\parameter) \right] 
        - \Pr_{W_{1} \gets \mu^\unif_{\Phi}} \left[ 1 \gets \cD^{W_1, W_1^\dag}(1^\secparam,1^\parameter) \right]} = \negl(\secparam).
    \end{align*}
    To ease notation, set $\Phi_0 = \Phi^*$ and $\Phi_1 = \Phi$. Then we have $\norm{\Phi_0 - \Phi_1}_\diamond = \norm{Q(\cdot) - \Phi(\cdot)}_\diamond \le \eps$. Take any $V_0 \in \cU(2^{m'})|_{\Phi_0}$ and $V_1 \in \cU(2^{m'})|_{\Phi_1}$. By \cref{def:unitary-extension} and \cref{thm:continuity-of-dilation}, there exists an $(m'-n')$-qubit unitary $U$ such that
    \begin{align*}
        \norm{(U \ot I_{n'}) V_0 (\ket{1^{m'-n}} \ot \cdot) - V_1 (\ket{1^{m'-n}} \ot \cdot) }_\op \le \sqrt{\norm{\Phi_0 - \Phi_1}_\diamond} = \sqrt{\eps}
    \end{align*}
    Hence, we have
    \begin{align*}
        \norm{(U \ot I_{n'}) V_0 (\ketbra{1^{m'-n}} \ot I_n) - V_1 (\ketbra{1^{m'-n}} \ot I_n) }_\op \le \sqrt{\eps}
    \end{align*}
    Define the projector $\Pi = \ketbra{1^{m'-n}} \ot I_n$. Then we have $\norm{V_1^\dag (U \ot I_{n'}) V_0 \Pi - \Pi }_\op \le \sqrt{\eps}$.

    \begin{claim}\label{claim:close-to-fixing-subspace}
        There exists a unitary operator $U' \in \cU(2^{m'})$ such that
    \begin{enumerate}
        \item $\norm{V_1^\dag (U \ot I_{n'}) V_0 - U'}_{\op} \le O(\sqrt{\eps})$.
        \item $U' \Pi = \Pi$.
    \end{enumerate}
    \end{claim}
    We will prove the claim after the theorem. Now we set $\wt{V}_0 = (U \ot I_{n'}) V_0$ and $\wt{V_1} = V_1 U'$. By the first item of \cref{claim:close-to-fixing-subspace}, we have $\norm{\wt{V_0} - \wt{V_1}}_{\op} \le O(\sqrt{\eps}) = \negl(\secparam)$. Also, we have
    \begin{align*}
        \Gamma_{\wt{V_0}}(\cdot) = \Tr_{[:-n']} \left[\wt{V_0} (1^{m'-n} \ot \cdot) \wt{V_0}^\dag\right] = \Tr_{[:-n']} \left[V_0 (1^{m'-n} \ot \cdot) V_0^\dag\right] = \Phi_0(\cdot)\\
        \Gamma_{\wt{V_1}}(\cdot) = \Tr_{[:-n']} \left[\wt{V_1} (1^{m'-n} \ot \cdot) \wt{V_1}^\dag\right] = \Tr_{[:-n']} \left[V_1 (1^{m'-n} \ot \cdot) V_1^\dag\right] = \Phi_1(\cdot)
    \end{align*}
    where the first equation uses the cyclic property of partial trace and the second equation uses $(1^{m'-n} \ot \cdot) = \Pi (1^{m'-n} \ot \cdot) \Pi^\dag$ and $U' \Pi = \Pi$ according to \cref{claim:close-to-fixing-subspace}. Therefore, we have $\wt{V_0} \in \cU(2^{m'})|_{\Phi_0}$ and $\wt{V_1} \in \cU(2^{m'})|_{\Phi_1}$. By \cref{rmk:sampling-uniform-distribution}, one can sample $W_b \gets \mu^\unif_{\Phi_b}$ by setting
    \begin{align*}
        W_b \gets (R \ot I_{n'}) \wt{V_b} R' \text{ ,\; where } R \sim \mu^\Haar_{2^{m'-n'}}\;, R' \sim \mu^{\Haar}_{\Span \set{\ket{x}}_{x \in [2^{m'}-2^n]}}.
    \end{align*}
    Therefore, for every polynomial-query distinguisher $\cD$, there is a polynomial-query distinguisher $\wt{\cD}$ that samples and applies $R,R'$ in forward and inverse directions, such that
    \begin{align*}
        \Pr_{W_b \gets \mu^\unif_{\Phi_b}}\left[ 1 \gets \cD^{W_b,W_b^\dag}(1^\secparam,1^\parameter) \right] = \Pr\left[ 1 \gets \wt{\cD}\;{}^{\wt{V_b},\wt{V_b}^\dag}(1^\secparam,1^\parameter) \right]
    \end{align*}
    for $b \in \zo$. Since $\norm{\wt{V_0} - \wt{V_1}}_{\op} \le \negl(\secparam)$ and thus $\norm{\wt{V_0}^\dag - \wt{V_1}^\dag}_{\op} \le \negl(\secparam)$, we obtain
    \begin{align*}
        &\abs{\Pr_{W_{0} \gets \mu^\unif_{\Phi_0}} \left[ 1 \gets \cD^{W_0, W_0^\dag}(1^\secparam,1^\parameter) \right] 
        - \Pr_{W_{1} \gets \mu^\unif_{\Phi_1}} \left[ 1 \gets \cD^{W_1, W_1^\dag}(1^\secparam,1^\parameter)\right]}\\
        =& \abs{\Pr\left[ 1 \gets \wt{\cD}\;{}^{\wt{V_0},\wt{V_0}^\dag}(1^\secparam,1^\parameter) \right] - \Pr\left[ 1 \gets \wt{\cD}\;{}^{\wt{V_1},\wt{V_1}^\dag}(1^\secparam,1^\parameter) \right]} = \negl(\secparam)
    \end{align*}
    for every polynomial-query distinguisher $\cD$ as desired. Hence, we have proved simulation security in the general case.
    
    Finally, we show that the ideal functionality $\Ideal_{\secparam,\parameter,\Phi}$ can be efficiently implemented with negligible error from the security parameter $1^\secparam$ and any quantum circuit with parameter $\parameter$ that computes $\Phi$ with negligible error.
    \begin{bAlgorithm}{Efficient implementation of  $\Ideal_{\secparam,\parameter,\Phi}$.}
    \label{algo:implement-ideal-functionality}
    \begin{itemize}
        \item Take as input $1^\secparam$ and a circuit $Q$ with parameter $\parameter = (n,n',m,s)$ that computes $\Phi$.
        \item Let $\PRO$ be the $\left(\Span\set{\ket{x}}_{x\in \zo^{\secparam+m} \setminus [d]}\right)$-preserving path-recording oracle  with $d = 2^{\secparam+m}-2^n$ from \cref{lemma:our-PRO}, and $\PRO'$ be the path-recording oracle on $(\secparam+m-n')$ qubits from \cref{lemma:PRO-from-MH25}.
        \item On initialization, initialize $\gsL \gets \ket{\emptyset},\gsR \gets \ket{\emptyset}, \gsLprime \gets \ket{\emptyset},\gsRprime \gets \ket{\emptyset}$, and output $(1^\secparam,1^\parameter)$.
        \item On query $(b \in \set{\pm 1},\rho_{\gsA} \in \Den{\secparam+m})$,
        \begin{itemize}
            \item Create a circuit for $\CTRL_{1^\secparam}\text{-}U_Q$ by replacing each unitary gate $G$ in $Q$ with $\CTRL_{1^\secparam}\text{-}G$.
            \item Partition register $\gsA$ into $\gsAone,\gsAtwo$ with $(\secparam+m-n'), n'$ qubits, respectively.
            \item If $b=1$, apply $(\PRO)_{\gsA,\gsL,\gsR}$, apply $(\CTRL_{1^\secparam}\text{-}U_Q)_{\gsA}$, and then apply $(\PRO')_{\gsAone,\gsLprime,\gsRprime}$.
            \item If $b=-1$, apply $(\PRO'^\dag)_{\gsAone,\gsLprime,\gsRprime}$, apply $(\CTRL_{1^\secparam}\text{-}U_Q^\dag)_{\gsA}$, and then apply $(\PRO^\dag)_{\gsA,\gsL,\gsR}$. 
            \item Output the state on register $\gsA$.
            \end{itemize}
        \end{itemize}
    \end{bAlgorithm}

Algorithm \ref{algo:implement-ideal-functionality} is efficient because the (subspace-preserving) path-recording oracles $\PRO,\PRO'$ are efficient by \cref{lemma:PRO-from-MH25,lemma:our-PRO}. Let us consider the following hybrid algorithms.
\begin{itemize}
    \item $\mathsf{Alg}_0$: This is Algorithm \ref{algo:implement-ideal-functionality}.
    \item $\mathsf{Alg}_1$: On initialization, sample $R \sim \mu^\Haar_{S^\perp},\; R' \sim \mu^\Haar_{2^{m'-n'}}$ and output $(1^\secparam,1^\parameter)$.
    \begin{itemize}
        \item On query $b=1$ and $\rho_\gsA \in \Den{\secparam+m}$, apply $(R' \ot I_{n'}) (\CTRL_{1^\secparam}\text{-}U_Q) R$ on register $\gsA$.
        \item On query $b=-1$ and $\rho_\gsA \in \Den{\secparam+m}$, apply $ R^\dag (\CTRL_{1^\secparam}\text{-}U_Q^\dag) (R'^\dag \ot I_{n'})$ on register $\gsA$.
    \end{itemize}
    \item $\mathsf{Alg}_2$: This is the ideal functionality $\Ideal_{\secparam,\parameter,\Phi^*}$ where $\Phi^*(\cdot) := Q(\cdot)$.
    \item $\mathsf{Alg}_3$: This is the ideal functionality $\Ideal_{\secparam,\parameter,\Phi}$.
\end{itemize}
By the security of $\PRO,\PRO'$ (\cref{lemma:PRO-from-MH25,lemma:our-PRO}), $\mathsf{Alg}_0$ is statistically indistinguishable with negligible error from $\mathsf{Alg}_1$ against any polynomial-query adversary. By the previous analysis of the distribution of $\widehat{Q} = (R' \ot I_{n'}) (\CTRL_{1^\secparam}\text{-}U_Q) R$, we have $\widehat{Q} \sim \mu^{\unif}_{\Phi^*}(2^{\secparam+m})$, and therefore $\mathsf{Alg}_1$ is equivalent to $\mathsf{Alg}_2$. We have shown that $\mathsf{Alg}_2 = \Ideal_{\secparam,\parameter,\Phi^*}$ and $\mathsf{Alg}_3 = \Ideal_{\secparam,\parameter,\Phi}$ are statistically indistinguishable with negligible error given that $\norm{\Phi^* - \Phi}_\diamond \le \negl(\secparam)$ when proving the robust version of simulation security. Putting these results together, Algorithm \ref{algo:implement-ideal-functionality} efficiently implements $\Ideal_{\secparam,\parameter,\Phi}$ with negligible error.
\end{proof}

\begin{claim}[Restating \cref{claim:close-to-fixing-subspace}]
    Let $V$ be a unitary operator and $\Pi$ be a projector such that $\norm{ V \Pi -\Pi}_{\op} \le \delta < \frac{1}{2}$. 
    Then there exists a unitary operator $W$ such that
    \begin{enumerate}
        \item $\norm{V-W}_{\op} \le O(\delta)$.
        \item $W \Pi = \Pi$.
    \end{enumerate}
\end{claim}
\begin{proof}
    Define another projector $\Pi' = V \Pi V^\dag$. Then
    \begin{align*}
        \norm{\Pi' - \Pi}_{\op} &\le \norm{V\Pi V^\dag - V \Pi}_{\op} + \norm{V \Pi - \Pi}_{\op}\\
        &\le \norm{\Pi V^\dag - \Pi}_{\op} + \norm{V \Pi - \Pi}_{\op}\\
        &\le \norm{(V \Pi - \Pi)^\dag}_{\op} + \norm{V \Pi - \Pi}_{\op} \le 2\delta
    \end{align*}
    By Jordan's lemma (\cref{lem:Jordan}), there exists a decomposition of the ambient Hilbert space into the direct sum $\bigoplus_{j} \cS_j$ of orthogonal subspaces $\set{\cS_j}_j$, where each $\cS_j$ has dimension at most $2$ and is invariant under $\Pi$ and $\Pi'$. Moreover, for $\cS_j$ of dimension $2$, there exist orthogonal bases $\set{\ket{u_{j}}, \ket{u_{j}^\perp}}$ and $\set{\ket{v_{j}}, \ket{v_{j}^\perp}}$ such that the restrictions of $\Pi$ and $\Pi'$ to the invariant subspace $\cS_j$ are $\ketbra{u_{j}}$ and $\ketbra{v_{j}}$, respectively. By a change of global phase, we can assume $\langle u_{j} \ket{v_{j}}, \langle u_{j} \ket{v_{j}^\perp}, \langle u_{j}^\perp \ket{v_{j}}, \langle u_{j}^\perp \ket{v_{j}^\perp}  \in \bbR_{\ge 0}$ without loss of generality. Let $\theta_j \in [0,\frac{\pi}{2}]$ be such that $\langle u_{j} \ket{v_{j}} = \cos \theta_j$. Then we have $\sin \theta_j = \norm{\ketbra{v_j} - \ketbra{u_j}}_\op \le \norm{\Pi' - \Pi}_\op \le 2 \delta$.
    Set $R = \bigoplus_j R_j$ where 
    \begin{itemize}
        \item $R_j$ is the identity operator on $\cS_j$ if the subspace has dimension $1$.
        \item $R_{j}$ is the rotation operator on $\cS_j$ that maps $\ket{v_j}\mapsto\ket{u_j}$ and  $\ket{v_j^\perp}\mapsto\ket{u_j^\perp}$ if the subspace has dimension $2$.
    \end{itemize}
    Then $R$ is a unitary operator by construction. 
    Moreover, on $\cS_j$ of dimension $2$, we have
    \begin{align*}
        R \Pi'\ket{v_j} &= R\ket{v_j} = \ket{u_j} = \Pi \ket{u_j} = \Pi R \ket{v_j}\\
        R \Pi'\ket{v_j^\perp} &= 0 = \Pi \ket{u_j^\perp} = \Pi R \ket{v_j^\perp}
    \end{align*}
    On $\cS_j = \Span \set{\ket{x}}$ of dimension $1$, we have $\Pi \ket{x} = \lambda \ket{x}$ and $\Pi' \ket{x} = \lambda' \ket{x}$ for $\lambda,\lambda' \in \zo$. Then $\abs{\lambda - \lambda'} = \norm{\Pi' \ket{x} - \Pi \ket{x}} \le 2\delta < 1$ and therefore $\lambda = \lambda'$. So we have
    $$R \Pi' \ket{x} = \lambda' \ket{x} = \lambda \ket{x} =  \Pi R \ket{x}$$
    Therefore, we have $R\Pi' = \Pi R$. Also, since $R$ is defined as a direct sum, we have
    \begin{align*}
        \norm{R - I}_{\op} \le \max_j \norm{R_j - I_j}_\op \le \max_{j: dim(\cS_j)=2} 2\sin\left({\theta_j}/{2}\right) \le \max_{j: dim(\cS_j)=2}\; \frac{\sin\left(\theta_j\right)}{\cos\left({\theta_j}/{2}\right)} \leq O(\delta)
    \end{align*}
    Set $U = RV$ and  $W = \Pi + U(I - \Pi)$. Then $U\Pi = RV \Pi = R\Pi' V = \Pi RV = \Pi U$. Therefore, the unitary $U$ is block-diagonal in $(\Pi, I-\Pi)$, so $W$ is also a unitary. By the definition of $W$, we have
    $W \Pi = \Pi $. Furthermore, we have
    \begin{align*}
        \norm{W - V}_{\op} &\le \norm{W - U}_{\op} + \norm{U - V}_{\op}\\
        &= \norm{\Pi - U \Pi}_{\op} +  \norm{RV - V}_{\op}\\
        &\le \norm{\Pi - V \Pi}_{\op} + \norm{V\Pi - RV \Pi}_{\op} + \norm{RV - V}_{\op}\\
        &\le \norm{\Pi - V \Pi}_{\op} + 2 \norm{R - I}_{\op} \le O(\delta).
    \end{align*}
\end{proof}

We make some final remarks on basing our result on the pseudorandom oracle model instead of the classical oracle model.
In \cite{bartusek2026new}, they showed that the ideal obfuscation scheme for classical circuits proposed in \cite{jain2023pseudorandom} is post-quantum secure in the quantumly accessible pseudorandom oracle model. We recall the definition of the quantumly accessible pseudorandom oracle model and their theorem.

\begin{dfn}[\cite{jain2023pseudorandom,bartusek2026new}, Quantumly accessible pseudorandom oracle model]
Let $F=\{\PRF_k\}_k$ be a family of keyed pseudorandom functions. 
The quantumly accessible pseudorandom oracle model for $F$ provides an interface, accessible to all parties, that internally maintains a uniformly random permutation 
$P:\{0,1\}^\secparam \to \{0,1\}^\secparam$ and supports quantum superposition queries.  
The interface offers two operations:
\begin{itemize}
    \item $\mathsf{PRO}(\Gen,k) \mapsto P(k)$;
    \item $\mathsf{PRO}(\Eval,h,x)\mapsto \PRF_{P^{-1}(h)}(x)$.
\end{itemize}
\end{dfn}

\begin{theorem}[\cite{bartusek2026new}]
\label{pqideal}
There exists a post-quantum
ideal obfuscation scheme for all classical circuits in the quantumly accessible pseudorandom oracle model assuming a post-quantum subexponential-secure one-way function and functional encryption.
\end{theorem}

By \cref{pqideal}, we can instantiate the classical oracle with post-quantum ideal obfuscation in the quantumly accessible pseudorandom oracle model, and therefore we have the following corollary.

\begin{cor}[Obfuscation of arbitrary quantum circuits in the pseudorandom oracle model]
    There exists an ideal obfuscation scheme for the class $\Qarb$ of arbitrary quantum circuits in the quantumly accessible pseudorandom oracle model satisfying \cref{def:ideal-obfuscation} under the ideal functionality $\set{\Ideal_{\secparam,\parameter,\Phi}}_{\secparam,\parameter,\Phi}$ proposed in \cref{sec:invariant-measure}, assuming the existence of a post-quantum subexponential-secure one-way function and functional encryption.
\end{cor}

\ifdefined\ShowAuthor
\section*{Acknowledgment}
The authors would like to thank Andrea Coladangelo, Fermi Ma, and Stefano Tessaro for useful discussions. The authors thank Robert Huang and Eli Goldin for highlighting that this work is among the early applications of PRUs. Miryam Huang is supported by NSF CAREER award 2141536. Er-Cheng Tang is supported by NSF NRT-QL award 2021540. This work was done while Miryam Huang and Er-Cheng Tang were visiting the Simons Institute for the Theory of Computing.
\fi

\ifdefined\LLNCS
\bibliographystyle{splncs04}
\else
\bibliographystyle{alpha}
\fi
\bibliography{references}

\end{document}